\documentclass[useAMS]{mn2e}
\usepackage{times}
\usepackage{graphicx}
\usepackage{epsfig}
\usepackage{amssymb}

\title[Orbital evolution of transient Uranian co-orbitals]
      {Comparative orbital evolution of transient Uranian co-orbitals: 
       exploring the role of ephemeral multibody mean motion resonances}

\author[C. de la Fuente Marcos and R. de la Fuente Marcos]
       {C.~de~la~Fuente~Marcos\thanks{E-mail: nbplanet@fis.ucm.es}
         and
        R. de la Fuente Marcos \\
        Universidad Complutense de Madrid,
        Ciudad Universitaria, E-28040 Madrid, Spain}
\date{Accepted 2014 April 9.
      Received 2014 April 7;
      in original form 2014 February 17}
\pubyear{2014}
\begin{document}
   \maketitle

   \begin{abstract}
      Uranus has three known co-orbitals: 83982~Crantor~(2002~GO$_{9}$), 
      2010~EU$_{65}$ and 2011~QF$_{99}$. All of them were captured in 
      their current resonant state relatively recently. Here, we perform 
      a comparative analysis of the orbital evolution of these transient 
      co-orbitals to understand better how they got captured in the 
      first place and what makes them dynamically unstable. We also look 
      for additional temporary Uranian co-orbital candidates among known 
      objects. Our $N$-body simulations show that the long-term 
      stability of 2011~QF$_{99}$ is controlled by Jupiter and Neptune; 
      it briefly enters the 1:7 mean motion resonance with Jupiter and 
      the 2:1 with Neptune before becoming a Trojan and prior to leaving 
      its tadpole orbit. During these ephemeral two-body mean motion 
      resonance episodes, apsidal corotation resonances are also 
      observed. For known co-orbitals, Saturn is the current source of 
      the main destabilizing force but this is not enough to eject a 
      minor body from the 1:1 commensurability with Uranus. These 
      objects must enter mean motion resonances with Jupiter and Neptune 
      in order to be captured or become passing Centaurs. Asteroid 
      2010~EU$_{65}$, a probable visitor from the Oort cloud, may have 
      been stable for several Myr due to its comparatively low 
      eccentricity. Additionally, we propose 2002~VG$_{131}$ as the 
      first transient quasi-satellite candidate of Uranus. Asteroid 
      1999~HD$_{12}$ may signal the edge of Uranus' co-orbital region. 
      Transient Uranian co-orbitals are often submitted to complex 
      multibody ephemeral mean motion resonances that trigger the 
      switching between resonant co-orbital states, making them 
      dynamically unstable. In addition, we show that the orbital 
      properties and discovery circumstances of known objects can be 
      used to outline a practical strategy by which additional Uranus' 
      co-orbitals may be found.
   \end{abstract}

   \begin{keywords}
      celestial mechanics --
      minor planets, asteroids: individual: 1999~HD$_{12}$ --
      minor planets, asteroids: individual: 83982~Crantor~(2002~GO$_{9}$) --
      minor planets, asteroids: individual: 2002~VG$_{131}$ --
      minor planets, asteroids: individual: 2010~EU$_{65}$ --
      minor planets, asteroids: individual: 2011~QF$_{99}$ --
      planets and satellites: individual: Uranus.
   \end{keywords}

   \section{Introduction}
      Besides bound companions or natural satellites, planets may have unbound companions or co-orbitals, i.e. objects trapped in a 1:1 mean 
      motion resonance with the planet. Co-orbital bodies are not only interesting curiosities found in the celestial mechanics studies,
      but also represent temporary reservoirs for certain objects as well as the key to understand the origin of retrograde outer satellites 
      of the giant planets and the accretional processes in the early Solar system (see e.g. Namouni, Christou \& Murray 1999). These 
      co-orbital bodies can be primordial, if they were captured in their present resonant state early in the history of the Solar system 
      and have remained dynamically stable for billions of years, or transient, if they were captured relatively recently.
      \hfil\par
      Several thousand minor bodies are known to be currently trapped in the 1:1 mean motion resonance with Jupiter and most of them may 
      have remained as such for Gyr (see e.g. Milani 1993; Jewitt, Trujillo \& Luu 2000; Morbidelli et al. 2005; Yoshida \& Nakamura 2005; 
      Robutel \& Gabern 2006; Robutel \& Bodossian 2009; Grav et al. 2011; Nesvorn\'y, Vokrouhlick\'y \& Morbidelli 2013). Neptune also 
      hosts a likely large population of co-orbitals both long-term stable (e.g. Kortenkamp, Malhotra \& Michtchenko 2004; Nesvorn\'y \& 
      Vokrouhlick\'y 2009; Sheppard \& Trujillo 2006, 2010) and transient (e.g. de la Fuente Marcos \& de la Fuente Marcos 2012b,c). In 
      sharp contrast, the number of asteroids currently engaged in co-orbital motion with Uranus appears to be rather small and its nature 
      comparatively ephemeral. Uranus has Centaurs temporarily trapped in other mean motion resonances as well (Masaki \& Kinoshita 2003; 
      Gallardo 2006, 2014).
      \hfil\par
      The theoretical possibility of finding primordial or transient Uranian co-orbitals, in particular Trojans, has been a subject of study 
      for almost three decades. In general, numerical simulations predict that Uranus may have retained a certain amount of its primordial 
      co-orbital minor planet population and also that short-term stable (for some Myr) co-orbitals may exist. Zhang \& Innanen (1988a,b) 
      and Innanen \& Mikkola (1989) found that some tadpole orbits may survive for 10 Myr. Mikkola \& Innanen (1992) described tadpole 
      orbits stable for at least 20 Myr and with libration amplitudes close to 100\degr; horseshoe orbiters were only stable for shorter 
      periods of time ($<$10 Myr). Holman \& Wisdom (1993) found similar results. Gomes (1998) explained the observational absence of 
      primordial Uranus Trojans as a side effect of planetary migration. Wiegert, Innanen \& Mikkola (2000) focused on quasi-satellites and 
      found that some test orbits were stable for up to 1 Gyr but only at low inclinations ($<$2\degr) and with eccentricities in the range 
      0.1--0.15. Nesvorn\'y \& Dones (2002) concluded that any existing Uranus' primordial Trojan population should have now been depleted 
      by a factor of 100. Gacka (2003) also found short-term stable tadpole orbits. In a very detailed study, Marzari, Tricarico \& Scholl 
      (2003) identified a real absence of Trojan stable orbits at low libration amplitudes and also that, among the Jovian planets, Uranus 
      has the lowest probability of hosting long-term stable Trojans. Horner \& Evans (2006) stated that Uranus appears not to be able to 
      efficiently capture objects into the 1:1 commensurability today even for short periods of time. Lykawka \& Horner (2010) found that 
      Uranus should have been able to capture and retain a significant population of Trojan objects from the primordial planetesimal disc by 
      the end of planetary migration. These authors concluded that originally the orbits of these objects should have had a wide range of 
      orbital eccentricities and inclinations.
      \hfil\par
      Predictions from numerical simulations appear to be generally consistent with the available observational evidence that transient 
      Uranian co-orbitals do exist but primordial ones may not. So far, Uranus has only three known co-orbitals: 83982 Crantor (2002 GO$_{9}$) 
      (Gallardo 2006; de la Fuente Marcos \& de la Fuente Marcos 2013b), 2010 EU$_{65}$ (de la Fuente Marcos \& de la Fuente Marcos 2013b) 
      and 2011~QF$_{99}$ (Alexandersen et al. 2013a; Alexandersen et al. 2013b). Due to its present short data-arc (85 d), 2010 EU$_{65}$ is 
      better described as a candidate. Asteroids Crantor and 2010 EU$_{65}$ follow horseshoe orbits in a frame of reference rotating with 
      Uranus and 2011~QF$_{99}$ is an L$_4$ Trojan; the three of them are only short-term stable and, therefore, they must be captured 
      objects. Most published studies focus on what makes a hypothetical primordial population of Uranian co-orbitals dynamically unstable, 
      leaving the question of how the transient ones got captured unanswered.
      \hfil\par
      In absence of observational bias, the lack of a sizeable present-day population of minor bodies trapped in the 1:1 commensurability 
      with Uranus is probably due to persistent perturbations by the other giant planets. Uranus' Trojans are affected by high-order 
      resonances with Saturn (Gallardo 2006). In the case of current horseshoe librators, Saturn also appears to be the main source of the 
      destabilizing force (de la Fuente Marcos \& de la Fuente Marcos 2013b). Dvorak, Bazs\'o \& Zhou (2010) studied the stability of 
      hypothetical primordial Uranus' Trojans and concluded that the Trojan regions are mostly unstable. For these authors, the orbital 
      inclination is the key parameter to characterize the stability of Uranus' Trojans; only the inclination intervals (0, 7)\degr, (9, 
      13)\degr, (31, 36)\degr and (38, 50)\degr appear to be stable (regions A, B, C and D, respectively, in Dvorak et al. 2010). The 
      existence of these islands of stability enables the presence of Trojan companions to Uranus. Asteroid 2011~QF$_{99}$ appears to 
      inhabit one of these stable areas in orbital parameter space as its current orbital inclination is nearly 11\degr but it is not a 
      primordial Uranus' Trojan.
      \hfil\par
      Here, we revisit the subjects of capture and stability of current Uranian co-orbitals, providing an independent assessment of the 1:1 
      commensurability of the newly found Uranus' Trojan, 2011~QF$_{99}$, studying its future orbital evolution and looking into its 
      dynamical past. Our numerical investigation is aimed at adding more pieces to the overall puzzle of the apparent scarcity of Uranian 
      co-orbitals as we explore the role of multibody ephemeral mean motion resonances. On the other hand, the comparative study of both 
      the dynamics of the few known co-orbitals and candidates, and their discovery circumstances, reveals important additional clues to 
      solve this puzzle. This paper is organized as follows. In Section 2, we briefly discuss the numerical model used in our $N$-body 
      simulations. The current status of 2011~QF$_{99}$ is reviewed in Section 3. The capture mechanism and stability of 2011~QF$_{99}$ are 
      further studied in Section 4. We revisit the cases of Crantor and 2010 EU$_{65}$ in Section 5. Section 6 introduces a few new Uranus' 
      co-orbital candidates. In Section 7, we discuss our results on the stability of current Uranian co-orbitals. We present a practical 
      guide to discover additional objects in Section 8. Section 9 summarizes our conclusions.

   \section{Numerical integrations}
      In order to perform a comparative analysis of the orbital evolution of known transient Uranian co-orbitals, we solve directly the 
      Newtonian equations of motion by means of the Hermite integration scheme described by Makino (1991) and implemented by Aarseth (2003). 
      The adequacy of this integration approach for Solar system dynamical studies has been extensively tested by the authors (de la Fuente 
      Marcos \& de la Fuente Marcos 2012a,b,c, 2013a,b). The standard version of this $N$-body sequential code is publicly available from 
      the IoA web site\footnote{http://www.ast.cam.ac.uk/$\sim$sverre/web/pages/nbody.htm}. 
      \hfil\par
      Our calculations include the perturbations by the eight major planets, the Moon, the barycentre of the Pluto-Charon system and the 
      three largest asteroids, (1) Ceres, (2) Pallas and (4) Vesta. For accurate initial positions and velocities, we used the latest 
      Heliocentric ecliptic Keplerian elements provided by the Jet Propulsion Laboratory (JPL) online Solar system data 
      service\footnote{http://ssd.jpl.nasa.gov/?planet\_pos} (Giorgini et al. 1996) and initial positions and velocities based on the DE405 
      planetary orbital ephemerides (Standish 1998) referred to the barycentre of the Solar system. The orbits were computed 1 Myr forward 
      and backward in time. Additional details can be found in de la Fuente Marcos \& de la Fuente Marcos (2012a). 
      \hfil\par
      In addition to the orbital calculations completed using the nominal elements in Tables \ref{elements0} and \ref{elements1}, we have 
      performed 50 control simulations with sets of orbital elements obtained from the nominal ones and the quoted uncertainties. These 
      control orbits follow a Gaussian distribution in the six-dimensional space of orbital elements and their orbital parameters are 
      compatible with the observations within the 3$\sigma$ uncertainties (see Tables \ref{elements0} and \ref{elements1}), reflecting 
      the observational incertitude in astrometry. In the integrations, the relative error in the total energy was always as low as 
      $2.0\times10^{-15}$ or lower after a simulated time of 1 Myr. The corresponding error in the total angular momentum is several 
      orders of magnitude smaller. In all the figures, and unless explicitly stated, $t$ = 0 coincides with the JD2456800.5 epoch. 

   \section{2011~QF$_{99}$: a review}
      Asteroid 2011~QF$_{99}$ was discovered on 2011 August 29 at $r$ = 22.6 mag by M. Alexandersen observing from Mauna Kea with the 
      Canada--France--Hawaii Telescope (CFHT; Alexandersen et al. 2013a). The observations were made as part of an ongoing survey with CFHT, 
      looking for trans-Neptunian objects (TNOs) and minor bodies located between the giant planets. However, they were not submitted to the 
      Minor Planet Center (MPC) until March 2013 (the discovery was announced on 2013 March 18; Alexandersen et al. 2013a). The object was 
      identified due to its relatively high proper motion with respect to the background stars. It is comparatively large with $H$ = 9.7 mag 
      or a diameter close to 60 km assuming a 5 per cent albedo. Its orbit is now reasonably well determined with 29 observations spanning a 
      data-arc of 419 d and it is characterized by $a$ = 19.13 au, $e$ = 0.18 and $i$ = 10\fdg81 (see Table \ref{elements0}). Using 
      these orbital parameters, 2011~QF$_{99}$ has been classified as a Centaur by both the MPC Database\footnote{http://www.minorplanetcenter.net/db\_search} 
      and the JPL Small-Body Database\footnote{http://ssd.jpl.nasa.gov/sbdb.cgi}. However, its period of revolution around the Sun, 83.69 yr 
      at present, is very close to that of Uranus, 84.03 yr. That makes 2011~QF$_{99}$ a very good candidate to inhabit Uranus' co-orbital 
      region. 
      \hfil\par
%
%
      \begin{table}
         \centering
         \fontsize{8}{11pt}\selectfont
         \tabcolsep 0.3truecm
         \caption{Heliocentric Keplerian orbital elements of 2011~QF$_{99}$ used in this research. Values include the 1$\sigma$ uncertainty. 
                  The orbit is based on 29 observations spanning a data-arc of 419 d (Epoch = JD2456800.5, 2014-May-23.0; J2000.0 ecliptic 
                  and equinox. Source: JPL Small-Body Database).
                 }
         \begin{tabular}{lcc}
            \hline
             Semimajor axis, $a$ (au)                          & = & 19.132$\pm$0.013       \\
             Eccentricity, $e$                                 & = & 0.178$\pm$0.002        \\
             Inclination, $i$ (\degr)                          & = & 10.807\,0$\pm$0.001\,4 \\
             Longitude of the ascending node, $\Omega$ (\degr) & = & 222.507$\pm$0.002      \\
             Argument of perihelion, $\omega$ (\degr)          & = & 286.8$\pm$0.3          \\
             Mean anomaly, $M$ (\degr)                         & = & 271.2$\pm$0.6          \\
             Perihelion, $q$ (au)                              & = & 15.73$\pm$0.03         \\
             Aphelion, $Q$ (au)                                & = & 22.54$\pm$0.02         \\
             Absolute magnitude, $H$ (mag)                     & = & 9.7                    \\
            \hline
         \end{tabular}
         \label{elements0}
      \end{table}
%
%
      Asteroid 2011~QF$_{99}$ was identified as a Trojan of Uranus by Alexandersen et al. (2013b). It was the first Uranus' Trojan observed 
      and identified as such. Uranus' Trojans share the semimajor axis of Uranus but they may have different eccentricity and inclination. 
      In a frame of reference rotating with Uranus, they move in the so-called tadpole orbits around the Lagrangian equilateral points L$_4$ 
      and L$_5$; L$_4$ is located on the orbit of Uranus at some 60$^{\circ}$ ahead or east of Uranus, and L$_5$ is some 60$^{\circ}$ west. 
      In other words, and for those objects, the relative mean longitude $\lambda_{\rm r} = \lambda - \lambda_{\rm Uranus}$ oscillates 
      around the values +60\degr (L$_4$) or -60\degr (or +300\degr, L$_5$), where $\lambda$ and $\lambda_{\rm Uranus}$ are the mean 
      longitudes of the Trojan and Uranus, respectively. The mean longitude of an object is given by $\lambda$ = $M$ + $\Omega$ + $\omega$, 
      where $M$ is the mean anomaly, $\Omega$ is the longitude of the ascending node and $\omega$ is the argument of perihelion. The two 
      other main co-orbital states are quasi-satellite ($\lambda_{\rm r}$ oscillates around 0\degr) and horseshoe ($\lambda_{\rm r}$ 
      oscillates around the Lagrangian point L$_3$ which is located on the orbit of Uranus but at 180\degr from the planet) librator (see 
      e.g. Mikkola et al. 2006; Murray \& Dermott 1999, respectively, for further details). For any given planet-minor body pair, 
      $\lambda_{\rm r}$ measures how far in orbital phase the minor body moves with respect to the planet. Under certain conditions, complex 
      orbits, hybrid of two (or more) elementary resonant states, are also possible. For example, a compound orbit between the Trojan and 
      quasi-satellite states is also called a large-amplitude Trojan when the libration amplitude is less than 180$^{\circ}$. Such an orbit 
      encloses L$_4$ (or L$_5$) and Uranus itself. Recurrent transitions between resonant states are possible for objects with both 
      significant eccentricity and inclination (Namouni et al. 1999; Namouni \& Murray 2000).
      \hfil\par
%
%
      \begin{figure}
        \centering
         \includegraphics[width=\linewidth]{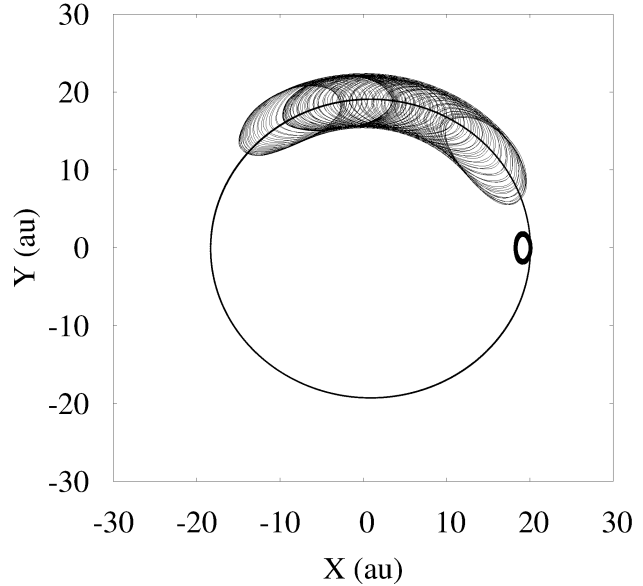}
         \caption{The motion of the asteroid 2011~QF$_{99}$ during the time interval (-5,~5) kyr projected on to the ecliptic plane in a 
                  coordinate system rotating with Uranus. The orbit and the position of Uranus are also plotted. In this frame of reference, 
                  and as a result of its non-negligible eccentricity, Uranus describes a small ellipse. This figure is equivalent to fig. 1 
                  in Alexandersen et al. (2013b). 
                 }
         \label{trojan}
      \end{figure}
\begin{figure}
        \centering
         \includegraphics[width=\linewidth]{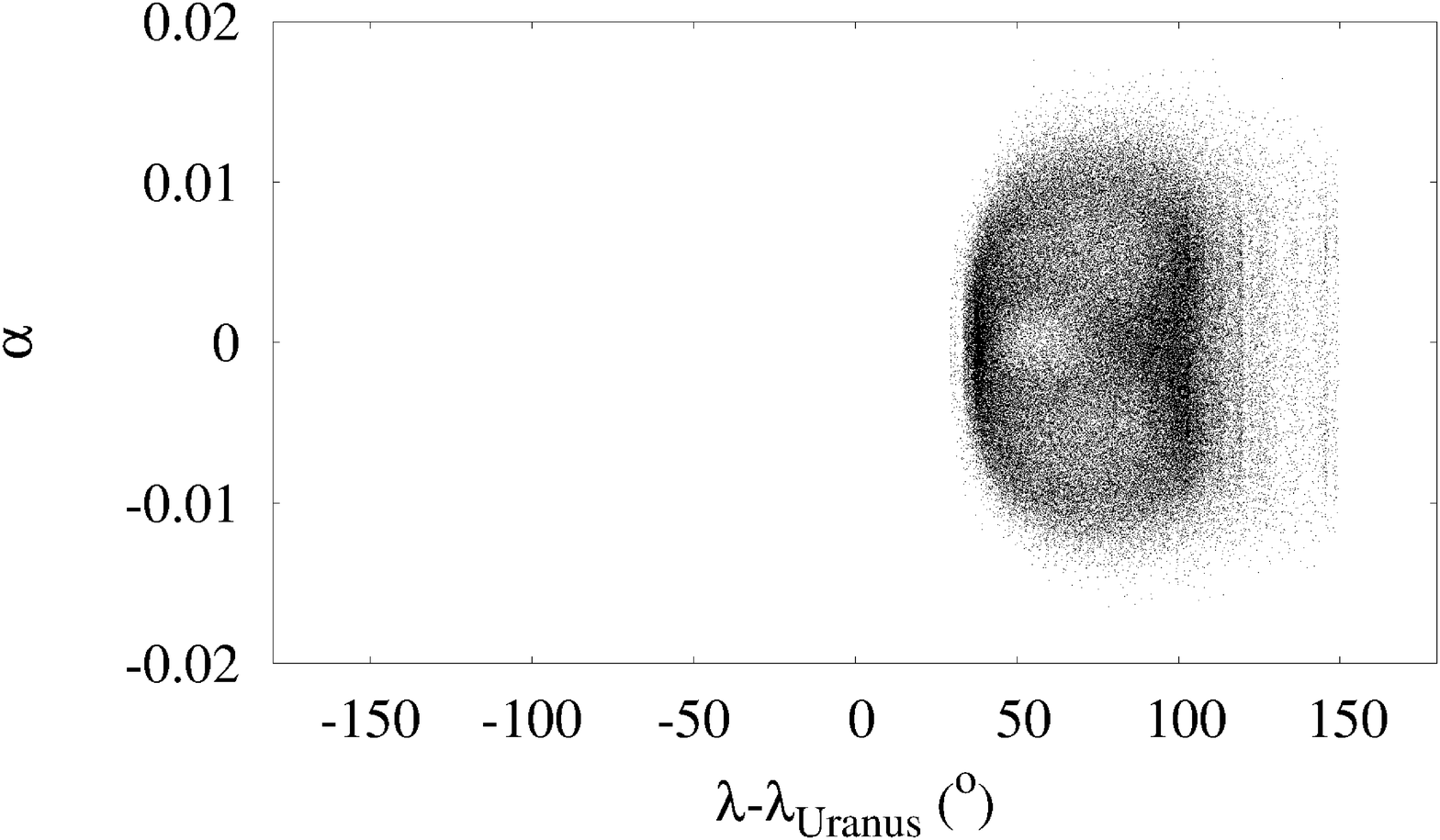}
         \caption{Resonant evolution of the asteroid 2011~QF$_{99}$ over the time range (-250, 250) kyr. The relative deviation of its 
                  semimajor axis from that of Uranus, $\alpha$, as a function of the relative mean longitude, $\lambda_{\rm r}$, is 
                  displayed.
                 }
         \label{resonant}
      \end{figure}

%
%
      In Fig. \ref{trojan}, we plot the motion of 2011~QF$_{99}$ over the time range (-5,~5) kyr in a coordinate system rotating with 
      Uranus. This figure is equivalent to fig. 1 in Alexandersen et al. (2013b). In this coordinate system, it appears to follow an 
      oscillating oval-like path (the epicycle) around Uranus' Lagrangian point L$_4$. Its $\lambda_{\rm r}$ librates around +60\degr with 
      an amplitude close to 100\degr. We therefore confirm the current dynamical status of 2011~QF$_{99}$ as Trojan. Following the work of 
      Mikkola et al. (2006), we define the relative deviation of the semimajor axis from that of Uranus by $\alpha = (a - a_{\rm Uranus}) / 
      a_{\rm Uranus}$, where $a$ and $a_{\rm Uranus}$ are the semimajor axes of the object and Uranus, respectively. If we plot $\alpha$ as 
      a function of $\lambda_{\rm r}$ over the time range (-250, 250) kyr, we obtain Fig. \ref{resonant} that further shows that the motion 
      of 2011~QF$_{99}$ is confined and also that the mean oscillation centre is very close to +60\degr in $\lambda_{\rm r}$. For a Trojan, 
      this object can stray relatively far from the Lagrangian point L$_4$ but it is not a large-amplitude Trojan because its orbit (in the 
      rotating frame) does not currently enclose Uranus. It is not a horseshoe-Trojan librator hybrid either, because the object does not 
      reach the Lagrangian point L$_3$. This minor planet has remained in its present dynamical state for hundreds of thousands of years and 
      it will continue to be trapped in this resonant configuration for some time into the future.
      \hfil\par
%
%
\begin{figure}
        \centering
         \includegraphics[width=\linewidth]{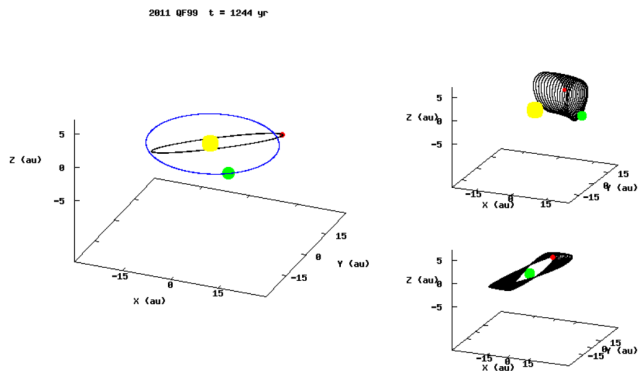}
         \caption{Three-dimensional evolution of the orbit of 2011~QF$_{99}$ in three different frames of reference: Heliocentric 
                  (left), frame corotating with Uranus but centred on the Sun (top-right), and Uranocentric (bottom-right). The 
                  red point represents 2011~QF$_{99}$, the green one is Uranus, and the yellow one is the Sun. The osculating 
                  orbits are outlined and the viewing angle changes slowly to facilitate visualizing the orbital evolution. This
                  figure is available on the electronic edition as a high-resolution animation.} 
        \label{animation}
     \end{figure}

%
%
      The current orbital behaviour of 2011~QF$_{99}$ is illustrated by the animation displayed in Fig. \ref{animation} (available on the 
      electronic edition as a high-resolution animation). The orbit is presented in three frames of reference: Heliocentric (left), 
      corotating with Uranus (top-right) and Uranocentric (bottom-right). The dynamical evolution of an object moving in a Trojan-like 
      orbit associated with Uranus can be decomposed into a slow guiding centre librational motion and a superimposed short period 
      three-dimensional epicyclic motion viewed in a frame of reference corotating with Uranus. The object spirals back and forth along 
      Uranus' orbit and ahead of the planet with a period of nearly 5500 yr. This period changes slightly as a result of the changes in 
      inclination. Each time the object gets close to Uranus, it is effectively repelled by the planet. In contrast, at the point of 
      maximum separation from Uranus it is drawn towards the planet. 
      \hfil\par
%
%
\begin{figure*}
        \centering
         \includegraphics[width=\linewidth]{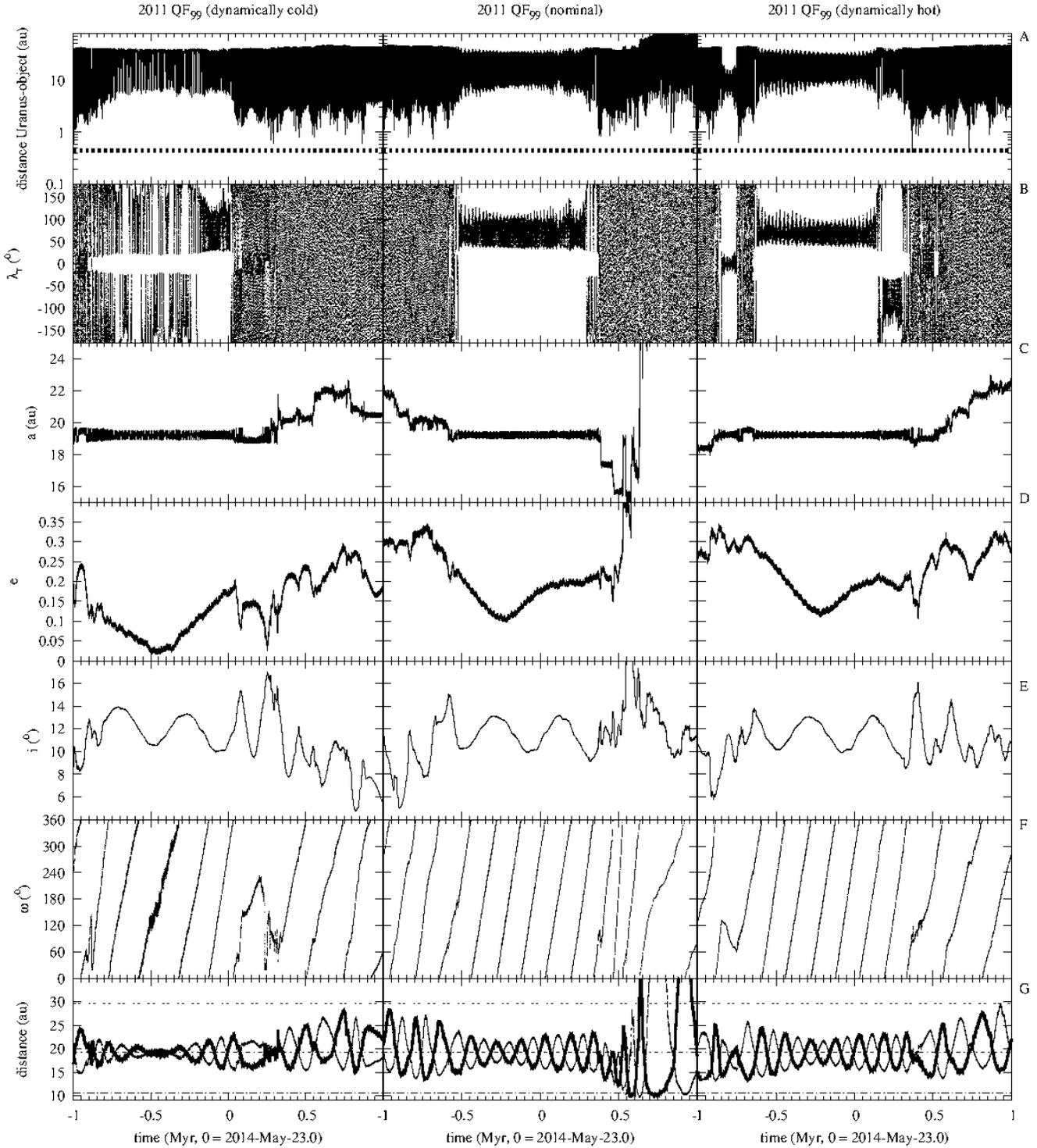}
         \caption{Time evolution of various parameters for the nominal orbit (central panels) and two representative examples of orbits that 
                  are most different from the nominal one in Table \ref{elements0} (see the text for details). The distance of 2011~QF$_{99}$ 
                  from Uranus (panel A); the value of the Hill sphere radius of Uranus, 0.447 au, is displayed. The resonant angle, 
                  $\lambda_{r}$ (panel B). The orbital elements $a$ (panel C) with the current value of Uranus' semimajor axis, $e$ (panel D), 
                  $i$ (panel E), and $\omega$ (panel F). The distances to the descending (thick line) and ascending nodes (dotted line) 
                  are shown in the G-panels. Saturn's and Neptune's aphelion and perihelion distances as well as Uranus semimajor axis are 
                  also plotted. Fig. S1 in Alexandersen et al. (2013b) is partially equivalent to this figure. 
                 }
         \label{all}
      \end{figure*}

%
%
      A plot of the evolution of the orbital elements of 2011~QF$_{99}$ over a 2 Myr interval centred on the present epoch is shown in Fig. 
      \ref{all}. This figure displays the dynamical evolution of the nominal orbit (central panels) and those of two representative worst 
      orbits which are most different from the nominal one. The orbit labelled as `dynamically cold' (left-hand panels) has been obtained by 
      subtracting three times the value of the respective uncertainty from the orbital parameters (the six elements) in Table \ref{elements0}. 
      This orbit is the coldest, dynamically speaking, possible (lowest values of $a$, $e$ and $i$) that is compatible with the current 
      values of the orbital parameters of 2011~QF$_{99}$. In contrast, the orbit labelled as `dynamically hot' (right-hand panels) was 
      computed by adding three times the value of the respective uncertainty to the orbital elements in Table \ref{elements0}. This makes 
      this trajectory the hottest possible in dynamical terms (largest values of $a$, $e$ and $i$). 
      \hfil\par
      All the control orbits exhibit consistent behaviour within a few dozen thousand years of $t = 0$. Dynamically colder orbits are less 
      stable. The distance to 2011~QF$_{99}$ from Uranus is plotted in Fig. \ref{all}, panel A and shows that, in the future, the object 
      undergoes close encounters with Uranus, near one Hill radius, prior to its departure from the co-orbital state. Such close encounters 
      are also observed in the past, before its insertion in the co-orbital region. The evolution of $\lambda_{r}$ in panel B indicates that, 
      for the nominal orbit, the current co-orbital episode will end in about 0.37 Myr from now; then, 2011~QF$_{99}$ will decouple from 
      Uranus with $\lambda_{r}$ stopping its current libration and starting to circulate. 
      \hfil\par
      The nominal orbit displays a fairly stable evolution during the entire co-orbital episode; in contrast, the slightly different (but 
      still compatible with the available observations) control orbits exhibit multiple transitions between the various co-orbital states. 
      This is similar to what is observed in the case of 83982 Crantor (2002 GO$_{9}$), a well studied Uranian horseshoe librator (de la 
      Fuente Marcos \& de la Fuente Marcos 2013b), even if this other object is far less stable (see below). Both L$_4$ and L$_5$ Trojan 
      episodes are possible but horseshoe-like behaviour and even quasi-satellite episodes are also observed. The semimajor axis, panel C, 
      exhibits the characteristic oscillatory behaviour associated with a 1:1 mean motion resonance. The value of the orbital eccentricity, 
      panel D, stays under 0.2 for most of the co-orbital evolution and the inclination, panel E, remains confined to region B as defined in 
      Dvorak et al. (2010). The argument of perihelion, panel F, mostly circulates, although some brief Kozai-like resonance (Kozai 1962) 
      episodes are observed (see Fig. \ref{all}, F-panels). 
      \hfil\par
      All control calculations indicate that 2011~QF$_{99}$ has been co-orbital with Uranus for at least 0.5 Myr but less than 2 Myr. It 
      will leave the co-orbital region in 50 kyr to 0.7 Myr from now. The most probable duration of the entire co-orbital episode is nearly 
      1 Myr. When trapped in the co-orbital region, its characteristic time-scale for chaotic divergence is $\sim$10 kyr. These results are 
      consistent with those in Alexandersen et al. (2013b). In the following section we further investigate the details of the mechanism 
      responsible for the insertion (ejection) of 2011~QF$_{99}$ into (from) the 1:1 mean motion resonance with Uranus.
%
%
%
%

   \section{2011~QF$_{99}$: stability analysis}
      Asteroid 2011~QF$_{99}$ is currently confined to region B as defined in Dvorak et al. (2010), inclination interval (9, 13)\degr. Our 
      numerical experiments indicate that, in this case, the instability surrounding this region is induced by the combined action of 
      Jupiter and Neptune. Dvorak et al. (2010) argued that the instability at the high-inclination boundary ($\sim$13\degr) is induced by 
      Jupiter and Uranus but they could not identify the cause of instability at the low-inclination boundary ($\sim$9\degr). Dvorak et al. 
      (2010) integrations only include the Sun and the four gas giants. They studied the semimajor axis range 18.9--19.6 au with the orbital 
      elements eccentricity, longitude of the node, and mean anomaly set to the corresponding values of Uranus ($e_{\rm Uranus}$ = 0.048, 
      $\Omega_{\rm Uranus}$ = 73\fdg84), and the argument of the perihelion $\omega = \omega_{\rm Uranus} \pm 60\degr$ ($\omega_{\rm Uranus}$ 
      = 95\fdg74). Some of these arbitrary choices have an impact on the applicability of their results to actual objects.
      \hfil\par 
      In any case, numerical integrations clearly indicate that the overall stability of individual Uranian co-orbitals not only depends on 
      the orbital inclination but also on the eccentricity. Uranus horseshoe librator 83982 Crantor (2002~GO$_{9}$) is also presently 
      confined to region B (even if it is not currently a Trojan) but its eccentricity is higher ($e$ = 0.27) and, consistently, the 
      destabilizing role of Saturn is enhanced with Neptune becoming a rather secondary perturber (de la Fuente Marcos \& de la Fuente 
      Marcos 2013b). The horseshoe librator candidate 2010~EU$_{65}$ is confined to a theoretically unstable region under Dvorak et al. 
      (2010) analysis for Trojans ($i$ = 14\fdg84) but its current dynamical status is significantly more stable than that of Crantor (or 
      even 2011~QF$_{99}$) due to its far lower eccentricity ($e$ = 0.05).  
      \hfil\par
\begin{figure}
        \centering
         \includegraphics[width=\linewidth]{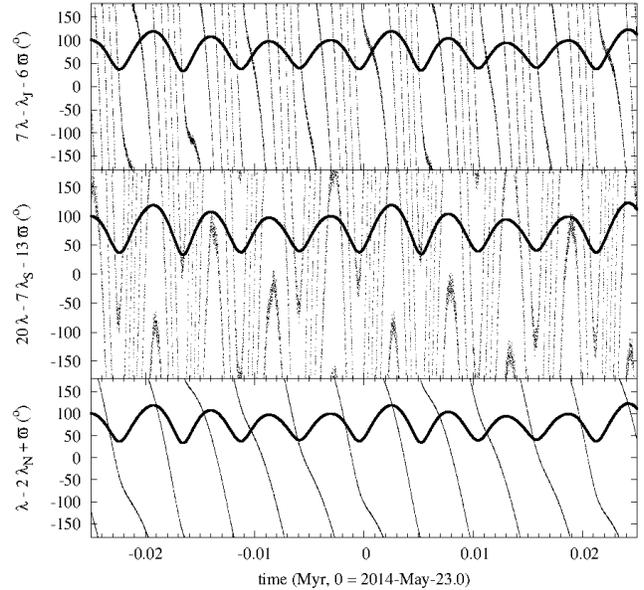}
         \caption{Asteroid 2011~QF$_{99}$. Resonant arguments $\sigma_{\rm J} = 7 \lambda - \lambda_{\rm J} - 6 \varpi$ (top panel), 
                  $\sigma_{\rm S} = 20 \lambda - 7 \lambda_{\rm S} - 13 \varpi$ (middle panel) and $\sigma_{\rm N} = \lambda - 2 \lambda_{\rm N} + 
                  \varpi$ (bottom panel) plotted against time for the time interval (-25, 25) kyr. The relative mean longitude with respect 
                  to Uranus appears as a thick line. The angle $\sigma_{\rm S}$ alternates between circulation and asymmetric libration, 
                  indicating that the motion is chaotic. The observed resonant evolution is consistent across control orbits. 
                 }
         \label{current}
      \end{figure}

      In its present orbit, 2011~QF$_{99}$ (and Uranus) moves in near resonance with the other three giant planets: 1:7 with Jupiter (period 
      ratio of 0.142), 7:20 with Saturn (period ratio of 0.351) and 2:1 with Neptune (period ratio of 1.963). The dominant role played by 
      Neptune and Jupiter on the orbital evolution of 2011~QF$_{99}$ must be connected with these near resonances and their possible 
      superposition. Multibody resonances are a major source of chaos in the Solar system. In order to understand better what makes 
      2011~QF$_{99}$ (and other Uranian co-orbitals) dynamically unstable we study the time evolution of the resonant arguments 
      $\sigma_{\rm J} = 7 \lambda - \lambda_{\rm J} - 6 \varpi$, $\sigma_{\rm S} = 20 \lambda - 7 \lambda_{\rm S} - 13 \varpi$ and 
      $\sigma_{\rm N} = \lambda - 2 \lambda_{\rm N} + \varpi$, where $\lambda_{\rm J}$ is the mean longitude of Jupiter, $\lambda_{\rm S}$ 
      is the mean longitude of Saturn, $\lambda_{\rm N}$ is the mean longitude of Neptune and $\varpi = \Omega + \omega$ is the longitude of 
      the perihelion of 2011~QF$_{99}$. In Fig. \ref{current}, arguments $\sigma_{\rm J}$, $\sigma_{\rm S}$, $\sigma_{\rm N}$ and the 
      relative mean longitude with respect to Uranus, $\lambda_{\rm r}$, (thick line) are plotted against time for the interval (-25, 25) 
      kyr. It is clear from the figure that present-day 2011~QF$_{99}$ is not in mean motion resonance with Jupiter or Neptune but it is 
      trapped in the 7:20 mean motion resonance with Saturn. This finding is consistent with the analysis carried out by Gallardo (2006) in 
      which he concluded that Uranus' Trojans are affected by high-order resonances with Saturn. The librations of both $\sigma_{\rm S}$ and 
      $\lambda_{\rm r}$ are synchronized and $\sigma_{\rm S}$ alternates between circulation and asymmetric libration, indicating that the 
      motion is chaotic. The argument $\sigma_{\rm J}$ exhibits a remarkable periodic behaviour but it does not librate; it may be very 
      close to the separatrix in the phase space of the mean motion resonance though. 
      \hfil\par
      For the nominal orbit, the resonant behaviour displayed in Fig. \ref{current} persists for about 0.7 Myr but it changes prior to the 
      insertion of 2011~QF$_{99}$ into Uranus' co-orbital region and also before its ejection from the 1:1 commensurability (see below). For 
      the control orbits, this resonant pattern is observed from 0.5 to over 2 Myr. As Trojan, 2011~QF$_{99}$ is submitted to a weak mean 
      motion resonance with Saturn, characterized by an intermittent asymmetric libration, that is unable to terminate its co-orbital state. 
      This is likely also the result of being very close to the separatrix in the phase space. In fact, this weak resonance with Saturn may 
      make the tadpole orbit more stable. Even if 2011~QF$_{99}$ is not strictly submitted to any mean motion resonance with Jupiter or 
      Neptune during its Trojan episode, their perturbations are strong enough to boost the libration amplitude of the tadpole path and 
      likely contribute towards the long-term instability (chaotic diffusion) of the present resonant configuration.   
      \hfil\par
%
%
\begin{figure}
        \centering
         \includegraphics[width=\linewidth]{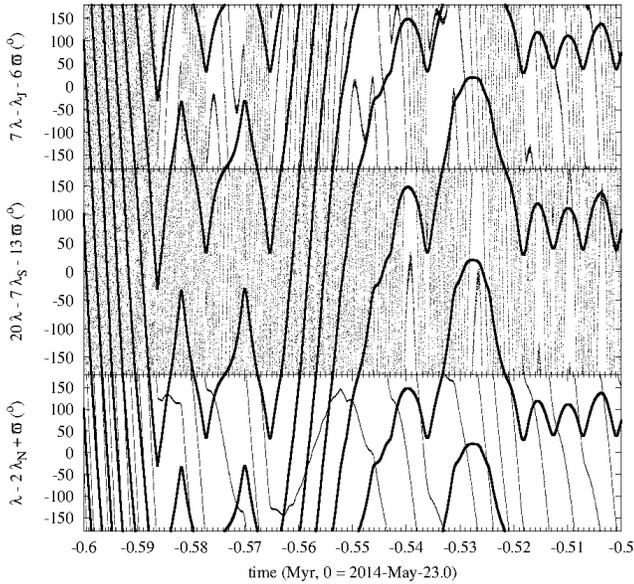}
         \caption{Asteroid 2011~QF$_{99}$. Similar to Fig. \ref{current} but for the time interval (-600, -500) kyr. The evolution of the 
                  various resonant arguments just prior to the insertion of 2011~QF$_{99}$ into Uranus' co-orbital region is displayed. 
                 }
         \label{insertion}
      \end{figure}

%
%
      Fig. \ref{insertion} shows the evolution of the various resonant arguments for the nominal orbit just prior to the insertion of 
      2011~QF$_{99}$ into Uranus' co-orbital region. All the resonant arguments plotted were circulating 590 kyr before the current epoch.
      From the point of view of Uranus, the asteroid was a passing object. Then, a close encounter with Uranus sent 2011~QF$_{99}$ into the 
      outskirts of Uranus' co-orbital region and into the 1:7 mean motion resonance with Jupiter and the 2:1 mean motion resonance with 
      Neptune. The combined action of these two mean motion resonances, a three-body mean motion resonance (Nesvorn\'y \& Morbidelli 1998; 
      Murray, Holman \& Potter 1998), made 2011~QF$_{99}$ a horseshoe librator first and later an L$_4$ Trojan. For over 50\,000 yr, the 
      three-body resonance was active, inserting the Centaur into a relatively stable configuration and also into the 7:20 mean motion 
      resonance with Saturn. Capture into Uranus' co-orbital zone takes place at the low-inclination boundary ($\sim$10\degr) of the 
      stability island described by Dvorak et al. (2010). Nearly 545 kyr ago, 2011~QF$_{99}$ left the three-body resonance with Jupiter and 
      Neptune and became firmly placed as Trojan. During the Trojan phase, 2011~QF$_{99}$ is also submitted to a three-body resonance, the 
      7:20 mean motion resonance with Saturn and the 1:1 mean motion resonance with Uranus. Remarkably, for the studied control orbits the 
      sequence of events leading to Trojan capture is, in general, almost identical to the one described for the nominal orbit, only the 
      timing is different.
      \hfil\par
%
%
\begin{figure}
        \centering
         \includegraphics[width=\linewidth]{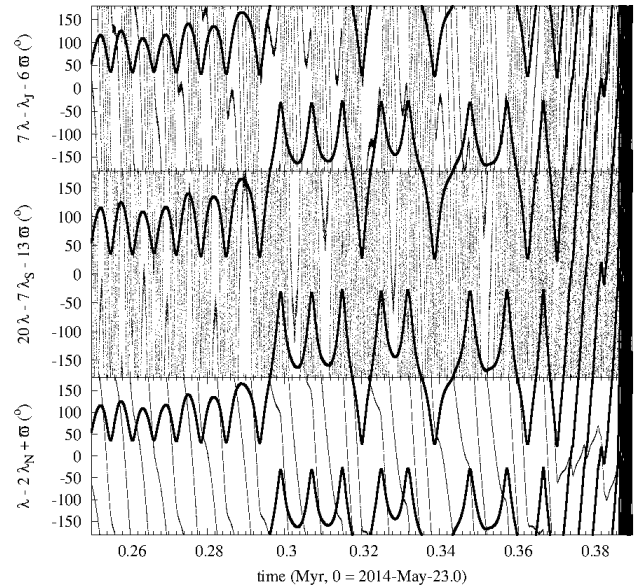}
         \caption{Asteroid 2011~QF$_{99}$. Similar to Fig. \ref{current} but for the time interval (250, 390) kyr. The evolution of the
                  various resonant arguments before becoming a passing object.
                 }
         \label{ejection}
      \end{figure}

%
%
      The ejection from Uranus' co-orbital region follows a similar process. Fig. \ref{ejection} shows the evolution of the same parameters 
      plotted in Figs \ref{current} and \ref{insertion} for the nominal orbit before 2011~QF$_{99}$ becomes a passing object. The object was 
      initially trapped in the 7:20 mean motion resonance with Saturn. Nearly 290 kyr into the future, the Trojan will fall again into the 
      1:7 mean motion resonance with Jupiter, leaving its tadpole orbit for an irregular horseshoe path. Nearly 80 kyr later, the object 
      will leave the resonance with Jupiter, briefly entering the 2:1 mean motion resonance with Neptune. This perturbation effectively 
      terminates the co-orbital episode with Uranus; the object becomes a passing Centaur again. Ejection from Uranus' co-orbital zone takes 
      place at the high-inclination boundary ($\sim$13\degr) of the stability island described by Dvorak et al. (2010). As in the case of 
      capture, the dynamical evolution of the control orbits during ejection events is similar to the one described for the nominal orbit 
      but with different timings. Our calculations clearly show that mean motion resonances with Jupiter and Neptune destabilize the 
      co-orbital dynamics of 2011~QF$_{99}$, eventually inducing close encounters with Uranus. This was first predicted by Marzari et al. 
      (2003).
      \hfil\par
      When studying isolated resonances, a critical angle is customarily defined so it exhibits different behaviour inside or outside the
      resonance, namely libration versus circulation. In contrast, our present study uncovers a resonant scenario characterized by highly
      nonlinear dynamics in which nested resonances are at work (see e.g. Morbidelli 2002). In addition, some of the resonances observed 
      here are ephemeral, i.e. comparatively short-lived. Under such conditions, it is difficult to define a critical angle (see e.g. 
      Robutel \& Gabern 2006). This is why we focus on the natural candidates for low-order mean motion resonances, neglecting the possible 
      existence of a, perhaps, better (and more complicated) critical angle relation that may be obtained by the means of an accurate 
      determination of the proper frequencies. For additional details on this interesting issue, see Robutel \& Gabern (2006).
      \hfil\par
      It could be argued that, in co-orbital regime, it is difficult to distinguish between the 2:1 mean motion resonance with Neptune and
      a secondary resonance. Secondary resonances were introduced by Lema\^{\i}tre \& Henrard (1990) and take place when there is a
      commensurability between apsidal and libration frequencies (for further details see Morbidelli 2002). Secondary resonances have been
      found for Jupiter's Trojan asteroids (see Robutel \& Gabern 2006). In our case, this is unlikely because the 2:1 resonance with 
      Neptune is only observed for a few thousand years, during capture and ejection, not during the relatively long trapping inside the 1:1 
      mean motion resonance with Uranus. Secondary resonances may be present during the co-orbital phase as in the case of Jupiter's Trojans
      but they probably have a very minor role for these transient objects.
      \hfil\par
%
%
%
%
      In general, minor bodies that cross the paths of one or more planets can be rapidly destabilized by scattering resulting from close 
      planetary approaches if their orbital inclinations are low. Asteroid 2011~QF$_{99}$ follows an eccentric orbit ($e\approx0.2$) but, 
      currently, it only crosses the orbit of Uranus. This is also the case of Crantor and 2010 EU$_{65}$ (de la Fuente Marcos \& de la 
      Fuente Marcos 2013b), and like them, only close encounters with Uranus can activate or deactivate their co-orbital status with Uranus. 
      All of them have similar, relatively significant orbital inclinations. In the Solar system, and for objects moving in inclined orbits, 
      close encounters with major planets are only possible in the vicinity of the nodes. The distance between the Sun and the nodes is 
      given by $r = a (1 - e^2) / (1 \pm e \cos \omega)$, where the `+' sign denotes the ascending node and the `-' sign the descending 
      node. 
      \hfil\par
      Fig.  \ref{all}, central G-panel shows the evolution of the distance to the nodes of 2011~QF$_{99}$ in the time range (-1, 1) Myr for
      the nominal orbit in Table \ref{elements0}. As in the case of Crantor and 2010 EU$_{65}$, the evolution of the orbital elements in 
      Fig. \ref{all} shows that, for 2011~QF$_{99}$, changes in $\omega$ dominate those in $a$ and $e$ and largely control the positions of 
      the nodes. The precession rate of the nodes is rather regular during co-orbital episodes but becomes chaotic when $\lambda_{\rm r}$ 
      circulates. As pointed out below, this precession rate is somewhat synchronized with those of Jupiter and Uranus itself. Although the 
      values of the nodal distances are currently very close to the value of the semimajor axis of Uranus, close (but relatively distant due 
      to its Trojan status) flybys are only possible at the ascending node because the object is always east of Uranus when confined to the 
      Trojan L$_4$ region. That makes it considerably more stable than Crantor, which experiences close encounters with Uranus at both 
      nodes. For 2011~QF$_{99}$, the combined action of Jupiter and Neptune drives the body out of the stable island and eventually 
      completely out of the co-orbital region. This happens when $\omega$ becomes 0\degr which also implies that its ascending node is now 
      located at the perihelion distance i.e. the closest possible to Saturn and Jupiter. At that time, the descending node is found at the 
      aphelion distance, the closest possible to Neptune. Under this arrangement, resonant perturbations are most effective as the perihelia 
      of the various objects involved approximately lie along the same direction. Then, after ejection from the co-orbital region, close 
      encounters with Saturn become possible as the nodal distances are able to reach Saturn's aphelion (see Fig. \ref{all}, G-panels). 
      \hfil\par
\begin{figure}
       \centering
        \includegraphics[width=\linewidth]{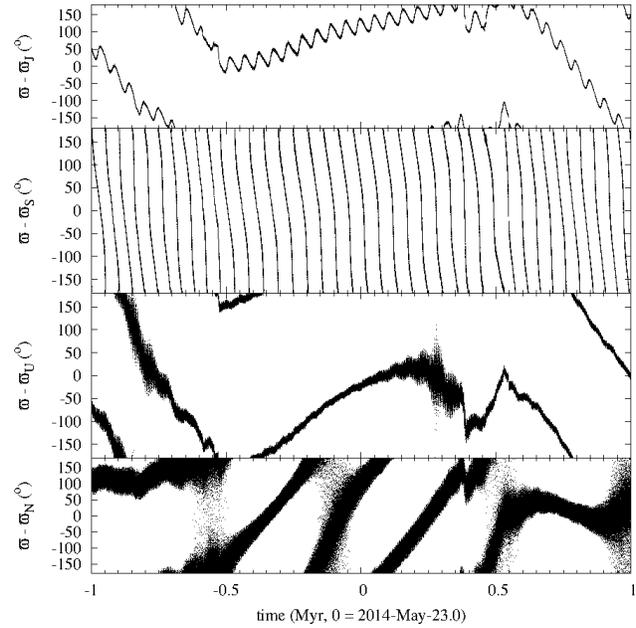}
        \caption{Time evolution of the relative longitude of the perihelion, $\Delta \varpi$, of 2011~QF$_{99}$ with respect to 
                 the giant planets: referred to Jupiter ($\varpi - \varpi_{\rm J}$), to Saturn ($\varpi - \varpi_{\rm S}$), to Uranus 
                 ($\varpi - \varpi_{\rm U}$) and to Neptune ($\varpi - \varpi_{\rm N}$). Only in the case of Saturn, the resonant 
                 argument $\Delta \varpi$ circulates over the entire simulated period. These results are for the nominal orbit in 
                 Table \ref{elements0}.
                }
        \label{rellon}
     \end{figure}

      The behaviour pointed out above suggests that the precession frequency of the longitude of the perihelion of 2011~QF$_{99}$, $\varpi = 
      \Omega + \omega$, could be in secular resonance with one or more of the giant planets. There is a well-documented secular resonance
      between Uranus' perihelion and Jupiter's aphelion, as the difference between the two librates around 130\degr within $\sim\pm$70\degr 
      with a period of $\sim$1.1 Myr (Milani \& Nobili 1985). This period is close to the most typical duration of the co-orbital episode of 
      2011~QF$_{99}$ with Uranus. Stockwell (1873) was first in realizing that the mean motion of Jupiter's perihelion is almost equal to 
      that of Uranus, and that the mean longitudes of these perihelia differ by nearly 180\degr. Following e.g. Laskar (1990), the frequency
      that dominates the precession of both Jupiter's and Uranus' perihelia is $g_5$. If we represent the relative longitude of the 
      perihelion, $\Delta \varpi = \varpi - \varpi_{\rm P}$, where $\varpi_{\rm P}$ is the longitude of the perihelion of the studied planet, 
      as a function of the time, we obtain Fig. \ref{rellon}. These plots show that only in the case of Saturn, the resonant argument 
      $\Delta \varpi$ circulates over the entire simulated time interval; the dominant frequency for Saturn's longitude of perihelion is 
      $g_6$ (see e.g. Laskar 1990). Before becoming a Uranian co-orbital both $\varpi - \varpi_{\rm J}$ and $\varpi - \varpi_{\rm U}$ 
      circulated. After leaving the Trojan region, $\varpi - \varpi_{\rm J}$ starts to circulate again. As predicted by Marzari et al. 
      (2003), during the Trojan episode the asteroid is in secular resonance with proper frequency $g_5$. 
      \hfil\par
      If $\Delta \varpi = \varpi - \varpi_{\rm P}$ librates around 0\degr, planet and object are very close to an stationary solution in 
      which both orbits have their semimajor axes aligned so the conjunction occurs when planet and object are both at perihelion. If the 
      libration is around 180\degr, then their perihelia lie in opposite directions. Fig. \ref{rellon} shows that, for a relatively brief 
      period of time, nearly 500 kyr ago, $\varpi - \varpi_{\rm J}$ librated symmetrically around 0\degr, when the object was also trapped 
      in the 1:7 mean motion resonance with Jupiter. During this brief period of time, the two bodies were in apsidal corotation resonance 
      (see Lee \& Peale 2002; Beaug\'e, Ferraz-Mello \& Michtchenko 2003) as their lines of apsides were aligned, maximizing the 
      perturbation. Simultaneously, $\varpi - \varpi_{\rm N}$ librated around 180\degr meaning that, for a brief period of time, their 
      perihelia lied on opposite directions. Prior to the ejection from the co-orbital region, $\varpi - \varpi_{\rm U}$ briefly librates 
      around 0\degr. Having their lines of apsides aligned, the two objects encounter each other at perihelion and that sends 2011~QF$_{99}$ 
      into the 1:7 mean motion resonance with Jupiter. Again an ephemeral double apsidal corotation is observed, symmetric with respect to 
      Jupiter and anti-symmetric with respect to Neptune, and the object becomes a passing Centaur. 
      \hfil\par
      The region where the motion of 2011~QF$_{99}$ occurs during co-orbital insertion and ejection is part of a very chaotic domain where 
      the superposition of mean motion and secular resonances severely complicates the dynamical study. Sometimes, the mean motion 
      resonances are dense in phase space but, in some cases, they actually overlap. The ephemeral episodes of apsidal corotation observed 
      between 2011~QF$_{99}$ and Jupiter corresponds to a Type I apsidal corotation as described in Beaug\'e et al. (2003).
%
%
%
%
%
%
     \begin{table*}
      \fontsize{8}{11pt}\selectfont
      \tabcolsep 0.2truecm
      \caption{Heliocentric Keplerian orbital elements (and the 1$\sigma$ uncertainty) of asteroids 1999~HD$_{12}$,
               83982 Crantor (2002~GO$_{9}$), 2002 VG$_{131}$ and 2010~EU$_{65}$ (Epoch = JD2456800.5, 2014-May-23.0; 
               J2000.0 ecliptic and equinox. Data for 1999~HD$_{12}$ are referred to epoch 2451300.5, 1999-May-02.0. 
               Data for 2002 VG$_{131}$ are referred to epoch 2452601.5, 2002-Nov-23.0. Source: JPL Small-Body Database.)
              }
      \begin{tabular}{lccccc}
       \hline
                                                          &   &   1999~HD$_{12}$ &   (83982) Crantor         &  2002 VG$_{131}$ &  2010~EU$_{65}$   \\
       \hline
        Semimajor axis, $a$ (au)                          & = &  19$\pm$32       &  19.357\,9$\pm$0.001\,4   &  19$\pm$12       &  19.3$\pm$0.3     \\
        Eccentricity, $e$                                 & = &   0.5$\pm$1.3    &   0.276\,03$\pm$0.000\,04 &   0.3$\pm$1.0    &   0.06$\pm$0.03   \\
        Inclination, $i$ (\degr)                          & = &  10.1$\pm$1.4    &  12.791\,26$\pm$0.000\,03 &  21$\pm$2        &  14.8$\pm$0.2     \\ 
        Longitude of the ascending node, $\Omega$ (\degr) & = & 178$\pm$2        & 117.403\,1$\pm$0.000\,3   & 213$\pm$2        &   4.583$\pm$0.011 \\ 
        Argument of perihelion, $\omega$ (\degr)          & = & 291$\pm$46       &  92.437$\pm$0.003         & 100$\pm$98       & 192$\pm$70        \\  
        Mean anomaly, $M$ (\degr)                         & = &  28$\pm$87       &  50.935$\pm$0.007         &  34$\pm$14       &   8$\pm$62        \\ 
        Perihelion, $q$ (au)                              & = &   9$\pm$9        &  14.014\,6$\pm$0.000\,2   &  13$\pm$10       &  18.1$\pm$0.3     \\ 
        Aphelion, $Q$ (au)                                & = &  29$\pm$48       &  24.701$\pm$0.002         &  26$\pm$17       &  20.5$\pm$0.3     \\ 
        Absolute magnitude, $H$ (mag)                     & = &  12.8            &   8.8                     &  11.3            &   9.1             \\
       \hline
      \end{tabular}
      \label{elements1}
     \end{table*}
%
%
%
%
      \begin{figure*}
        \centering
         \includegraphics[width=\linewidth]{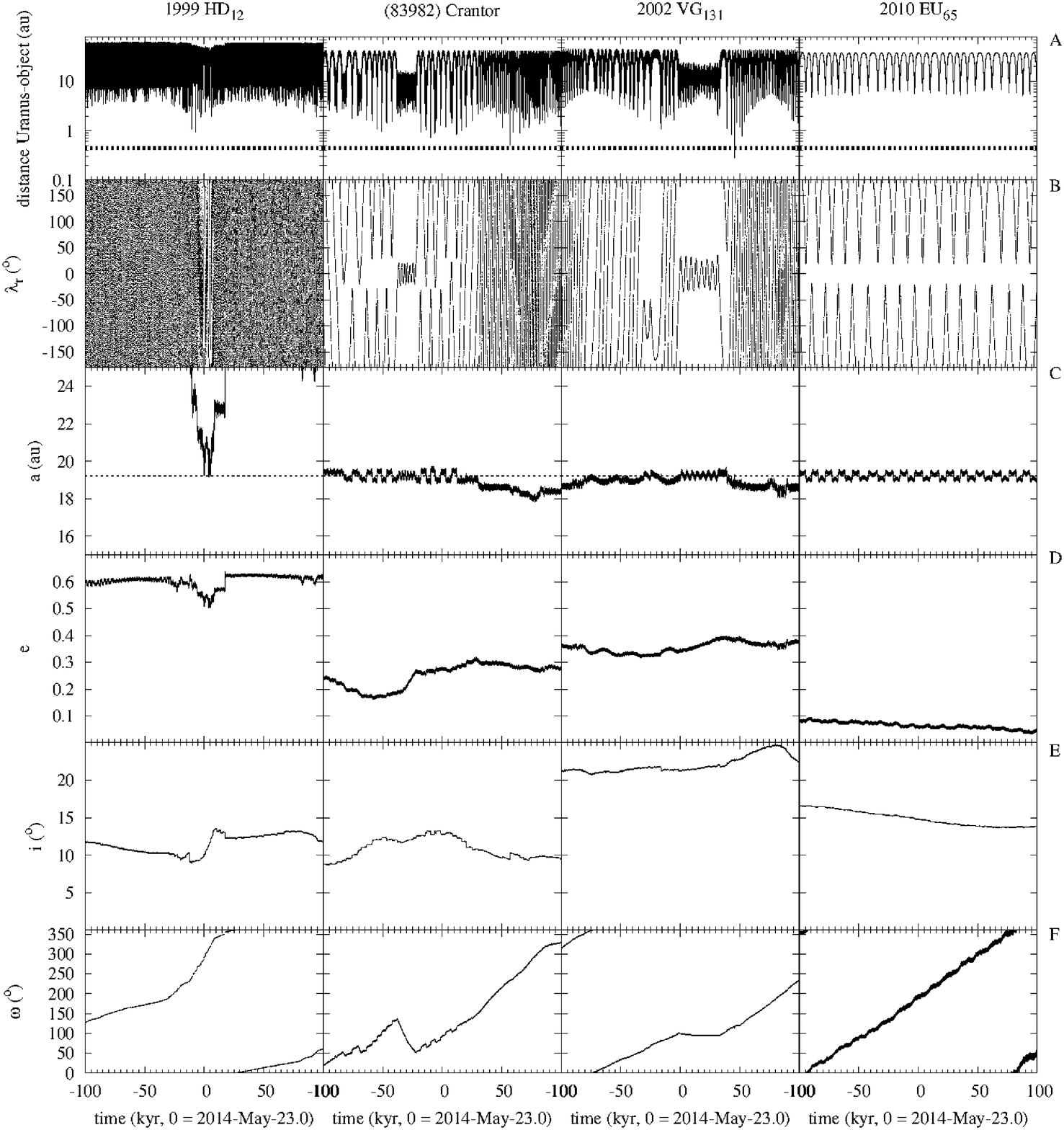}
         \caption{Time evolution of various parameters for the nominal orbits of known and new Uranian co-orbital candidates (see 
                  the text for details). The distance of the object from Uranus (panel A); the value of the Hill sphere radius of 
                  Uranus, 0.447 au, is displayed. The resonant angle, $\lambda_{\rm r}$ (panel B). The orbital elements $a$ (panel C) 
                  with the current value of Uranus' semimajor axis, $e$ (panel D), $i$ (panel E), and $\omega$ (panel F). 
                 }
         \label{others}
      \end{figure*}
%
%

   \section{Two horseshoe librators: 83982 Crantor (2002 GO$_{9}$) and 2010 EU$_{65}$}
\begin{figure}
        \centering
         \includegraphics[width=\linewidth]{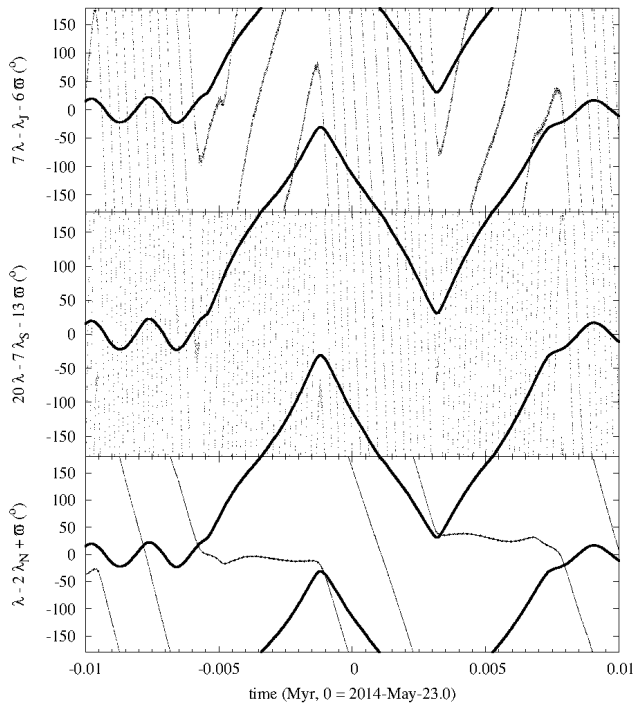}
         \caption{Asteroid 83982 Crantor (2002 GO$_{9}$). Resonant arguments $\sigma_{\rm J} = 7 \lambda - \lambda_{\rm J} - 6 \varpi$ 
                  (top panel), $\sigma_{\rm S} = 20 \lambda - 7 \lambda_{\rm S} - 13 \varpi$ (middle panel) and $\sigma_{\rm N} = \lambda - 
                  2 \lambda_{\rm N} + \varpi$ (bottom panel) plotted against time for the time interval (-10, 10) kyr. The relative mean 
                  longitude with respect to Uranus appears as a thick line. The resonant arguments exhibit alternating libration and 
                  circulation that trigger the transitions between the various co-orbital states. 
                 }
         \label{crantorR}
      \end{figure}

      Asteroid 83982 Crantor (2002 GO$_{9}$) was initially proposed as Uranus' first co-orbital companion by Gallardo (2006). In his work,
      Crantor is identified as horseshoe librator. Using an improved orbit, de la Fuente Marcos \& de la Fuente Marcos (2013b) confirmed 
      that Crantor is, in fact, a temporary horseshoe librator. The short-term evolution of Crantor is displayed in Fig. \ref{others} and
      it is substantially less stable than that of 2011~QF$_{99}$ in Fig. \ref{all}. Fig. \ref{crantorR} shows that the orbital evolution of 
      this object is alternatively affected by the 1:7 mean motion resonance with Jupiter, the 7:20 with Saturn and the 2:1 with Neptune.  
      At the present time, it is not actively submitted to any of these resonances but it was the combined action of the 1:7 mean motion 
      resonance with Jupiter and the 2:1 mean motion resonance with Neptune, the one responsible for sending Crantor from the 
      quasi-satellite to the horseshoe state nearly 6 kyr ago. The combination of resonances with Jupiter and Neptune will send the
      object back to the quasi-satellite state in about 7 kyr. During horseshoe librator episodes the relative mean longitude with respect 
      to Uranus librates in synchrony with the resonant argument $\sigma_{\rm S}$; when Crantor is closest to Uranus, $\sigma_{\rm S}$ 
      librates around 180\degr. Control orbits exhibit almost identical behaviour during horseshoe episodes (just slight difference in 
      timings) but in most cases the 7:20 mean motion resonance with Saturn is also observed during the quasi-satellite phase. Differences
      observed between orbits starting with very similar initial conditions ($<$1$\sigma$ difference) are the result of chaos.
      \hfil\par
%
%
\begin{figure}
        \centering
         \includegraphics[width=\linewidth]{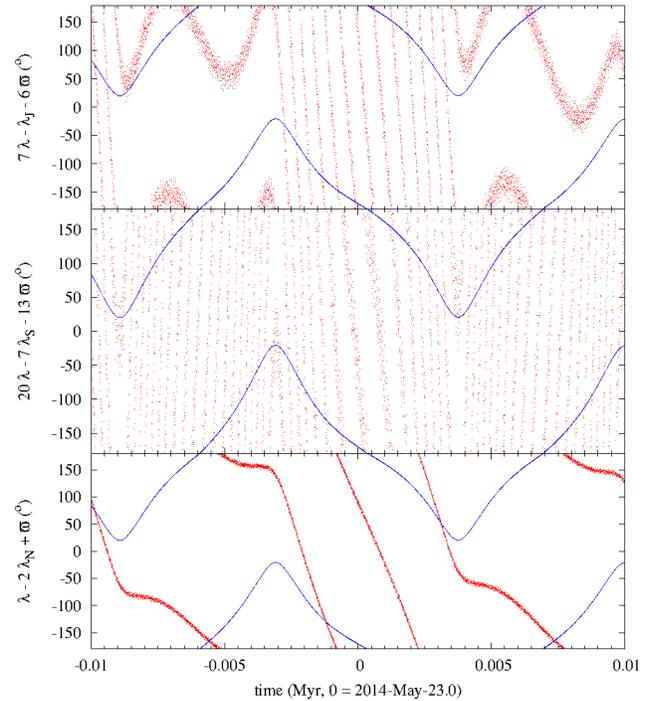}
         \caption{Asteroid 2010 EU$_{65}$. Resonant arguments $\sigma_{\rm J} = 7 \lambda - \lambda_{\rm J} - 6 \varpi$ (top panel), 
                  $\sigma_{\rm S} = 20 \lambda - 7 \lambda_{\rm S} - 13 \varpi$ (middle panel) and $\sigma_{\rm N} = \lambda - 
                  2 \lambda_{\rm N} + \varpi$ (bottom panel) plotted against time in red for the time interval (-10, 10) kyr. The relative 
                  mean longitude with respect to Uranus appears in blue. Both $\sigma_{\rm J}$ and $\sigma_{\rm S}$ alternate between 
                  circulation and asymmetric libration, indicating that the motion is chaotic but the resonant argument $\sigma_{\rm S}$ is 
                  synchronized with $\lambda_{\rm r}$. The argument $\sigma_{\rm N}$ mostly circulates.
                 }
         \label{eu65R}
      \end{figure}

\begin{figure}
        \centering
         \includegraphics[width=\linewidth]{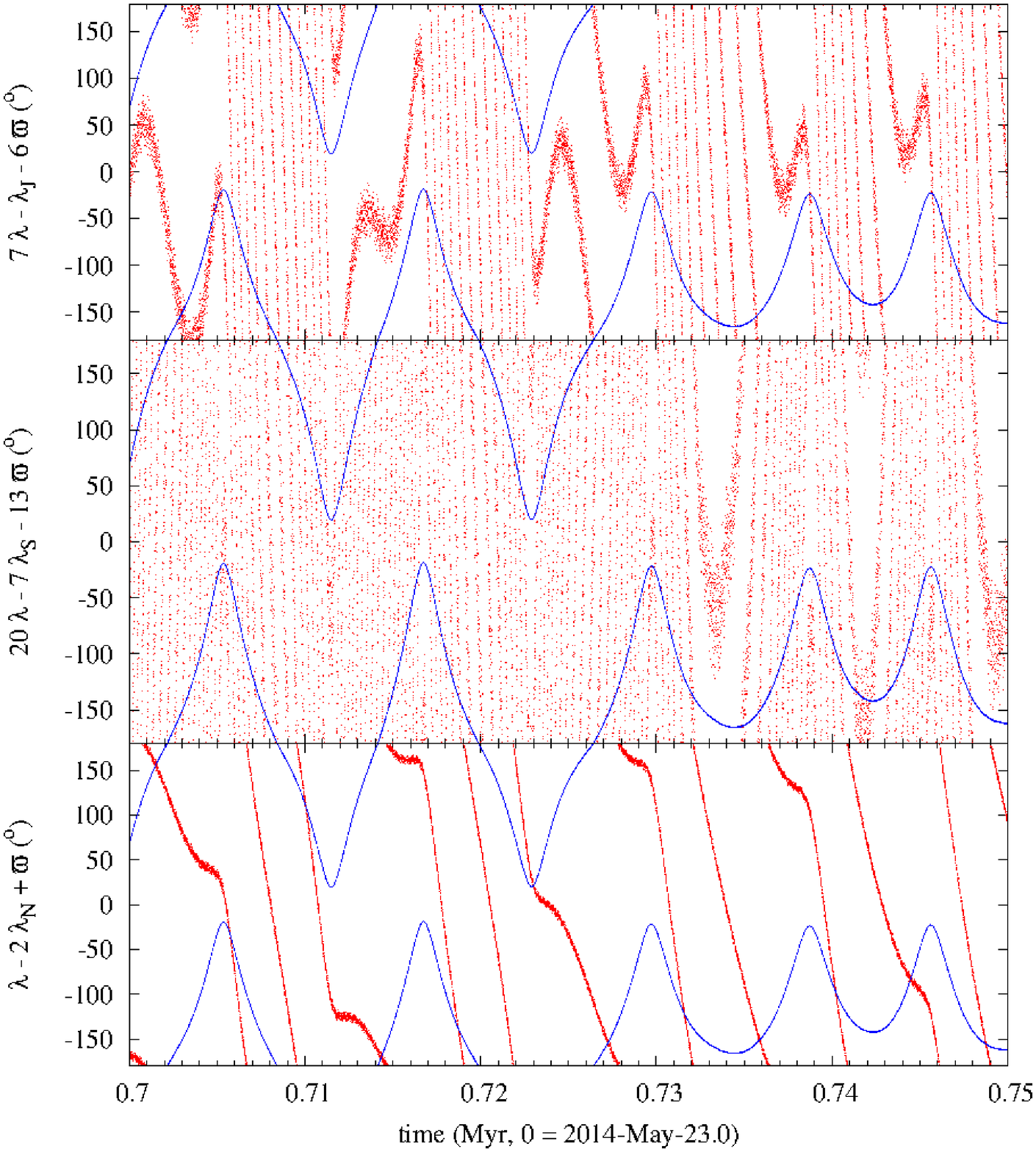}
         \caption{Same as Fig. \ref{eu65R} but for the time interval (0.70, 0.75) Myr. 
                 }
         \label{eu65Rw}
      \end{figure}

%
%
      Objects moving in orbits similar to that of 2010 EU$_{65}$ are very stable Uranian co-orbitals, significantly more stable than 
      2011~QF$_{99}$. All the control orbits indicate that this object may remain as a co-orbital companion to Uranus alternating the Trojan
      and horseshoe librator states for several ($>$ 2) Myr. If we study the resonant arguments $\sigma_{\rm J}$, $\sigma_{\rm S}$ and 
      $\sigma_{\rm N}$ (see Figs \ref{eu65R} and \ref{eu65Rw}), we observe that this object is also trapped in a mean motion resonance with 
      Saturn during its co-orbital evolution as the other objects are at some point. The relative mean longitude with respect to Uranus 
      librates in synchrony with the resonant argument $\sigma_{\rm S}$. Both $\sigma_{\rm J}$ and $\sigma_{\rm S}$ alternate between 
      circulation and asymmetric libration, indicating that the motion is chaotic. The argument $\sigma_{\rm N}$ mostly circulates but it 
      must be very close to the separatrix. All the control orbits indicate that 2010 EU$_{65}$ is a relatively recent visitor probably from 
      the Oort cloud. The object was likely captured in the 1:1 resonance with Uranus 1--3 Myr ago. If originated in the Oort cloud, it 
      may have entered the Solar system nearly 3 Myr ago. It will remain as Uranus' co-orbital for 2--5 more Myr. After leaving Uranus' 
      co-orbital region it could remain between the orbits of Uranus and Neptune for some time but it may also be ejected from the Solar 
      system. Its co-orbital episode with Uranus will last for about 3--8 Myr.
%
%
%
%
%
%
%
%

   \section{Uranian co-orbitals: are there others?}
      Asteroid 2011~QF$_{99}$ was found during a survey mainly focused on TNOs. During the past 15 years or so, wide-field CCD surveys aimed 
      at the outer Solar system have found dozens of objects moving in the neighbourhood of Uranus, many of which were classified using cuts 
      in the perihelion, $q$, and other orbital parameters (as an example, the MPC defines that Centaurs must have a perihelion larger than 
      Jupiter's orbit and a semimajor axis shorter than Neptune's). This approach leads to misidentification of resonant minor bodies 
      (Shankman 2012). Therefore, it is quite possible that some of these objects may have not been properly classified and that they may be 
      trapped, perhaps just temporarily, in a 1:1 mean motion resonance with Uranus. Here, we explore this possibility and try to uncover 
      additional Uranus' co-orbitals. We select candidates with relative semimajor axis, $|a - a_{\rm Uranus}| \leq$ 0.15 au, and use 
      $N$-body simulations, as described above, to confirm or reject their current co-orbital nature with Uranus. Our selection criterion 
      reveals two additional candidates: 1999~HD$_{12}$ and 2002~VG$_{131}$. We call them candidates because both objects have very short 
      data-arcs. As a result, their orbits are very poorly determined. They are included here for completeness, to explore the dynamics of 
      other possible short-term stable co-orbital parameter domains, and to encourage the acquisition of further observations to improve 
      their orbital solutions.
      \hfil\par
\begin{figure}
        \centering
         \includegraphics[width=\linewidth]{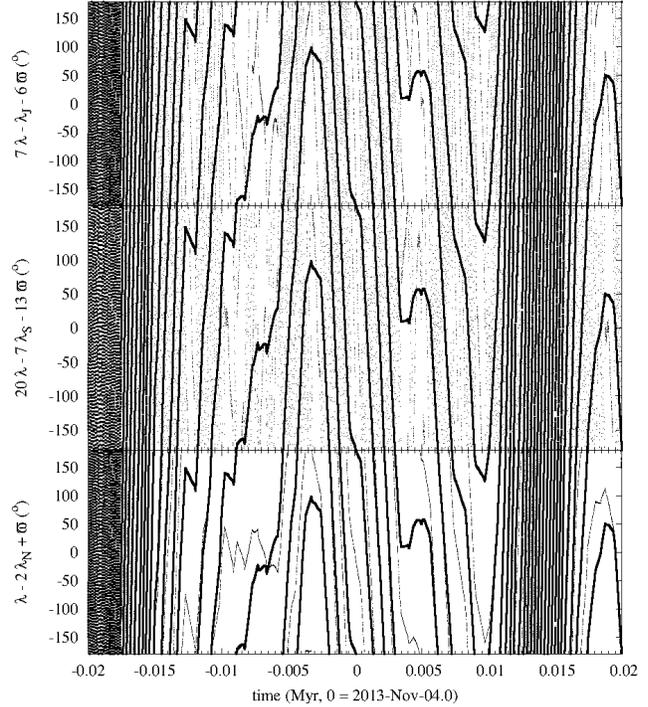}
         \caption{Asteroid 1999~HD$_{12}$. Resonant arguments $\sigma_{\rm J} = 7 \lambda - \lambda_{\rm J} - 6 \varpi$ (top panel), 
                  $\sigma_{\rm S} = 20 \lambda - 7 \lambda_{\rm S} - 13 \varpi$ (middle panel) and $\sigma_{\rm N} = \lambda - 
                  2 \lambda_{\rm N} + \varpi$ (bottom panel) plotted against time for the time interval (-20, 20) kyr. The relative mean 
                  longitude with respect to Uranus appears as a thick line. This object is submitted to a very complex set of mean motion 
                  resonances that involves all the giant planets.
                 }
         \label{hd12R}
      \end{figure}

      Asteroid 1999~HD$_{12}$ was discovered on 1999 April 17 at $R$ = 22.9 mag by the Deep Ecliptic Survey (DES) observing with the 4 m 
      Mayall Telescope at Kitt Peak National Observatory (Millis et al. 1999; Bernstein \& Khushalani 2000; Millis et al. 2002; Elliot et 
      al. 2005). The uncertainties associated with its orbit are very large as it is based on just 11 observations with a data-arc of only 49 
      d. The orbital elements are: $a$ = 19.34 au, $e$ = 0.52, $i$ = 10\degr, $\Omega$ = 178\degr and $\omega$ = 291\degr, with $H$ = 12.8 
      mag. Its orbital evolution is shown in Fig. \ref{others} and it is extremely chaotic. If 83982 Crantor (2002~GO$_{9}$) has an 
      $e$-folding time of nearly 1 kyr, any object moving in an orbit similar to that of 1999~HD$_{12}$ has a much shorter $e$-folding time 
      ($<$ 100 yr). With a perihelion of 9.4 au and an aphelion of 29.3 au, close encounters with Saturn, Uranus and Neptune are possible. 
      Its co-orbital episodes with Uranus are brief (a few kyr) and well scattered. Objects like this one may well signal the edge of Uranus' 
      co-orbital region. Fig. \ref{hd12R} shows that it is submitted to a very complex set of mean motion resonances that involves all the 
      giant planets. This explains the extremely chaotic behaviour observed (compare the evolution of $\lambda_{\rm r}$ for JD2456800.5 in 
      Fig. \ref{others} with that of JD2456600.5 in Fig. \ref{hd12R}, for the same input observations).  
%
%
%
%
      \hfil\par
\begin{figure}
        \centering
         \includegraphics[width=\linewidth]{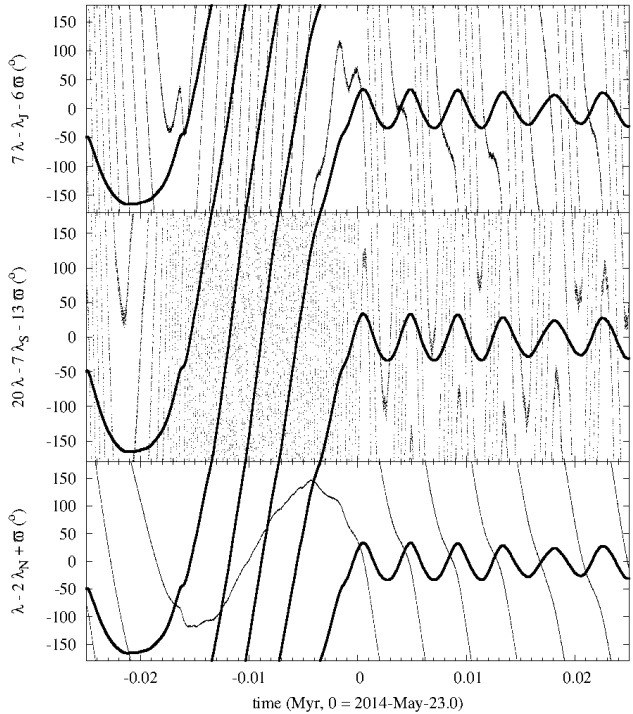}
         \caption{Asteroid 2002~VG$_{131}$. Resonant arguments $\sigma_{\rm J} = 7 \lambda - \lambda_{\rm J} - 6 \varpi$ (top panel), 
                  $\sigma_{\rm S} = 20 \lambda - 7 \lambda_{\rm S} - 13 \varpi$ (middle panel) and $\sigma_{\rm N} = \lambda - 
                  2 \lambda_{\rm N} + \varpi$ (bottom panel) plotted against time for the time interval (-25, 25) kyr. The relative mean 
                  longitude with respect to Uranus appears as a thick line. The resonant behaviour is very similar to that plotted in Fig. 
                  \ref{crantorR}.
                 }
         \label{vg131R}
      \end{figure}

      Asteroid 2002~VG$_{131}$ was discovered on 2002 November 9 at $R$ = 22.5 mag also by DES (Meech et al. 2002; Elliot et al. 2005). Its 
      orbit, based on six observations and spanning a data-arc of only 26 d, is very uncertain too. The orbital elements are: $a$ = 19.10 au, 
      $e$ = 0.34, $i$ = 21\degr, $\Omega$ = 213\degr and $\omega$ = 100\degr, with $H$ = 11.2 mag. It is currently a quasi-satellite, the 
      first identified candidate in this resonant state for Uranus. The orbital evolution of 2002~VG$_{131}$ as depicted in Fig. \ref{others} 
      closely resembles that of Crantor, with a very similar $e$-folding time. It is very likely that most present-day, transient Uranian 
      co-orbitals are dynamical analogues of Crantor and 2002~VG$_{131}$. This object has been included by Jewitt (2009) in his study of 
      active Centaurs with an effective radius of the nucleus of 11 km and a coma magnitude $>$26.13 mag (however, it was not active in 
      2002 December when it was observed by Keck). Regarding the influence of multibody resonances on the orbital evolution of this object,
      Fig. \ref{vg131R} shows that 2002~VG$_{131}$ is submitted to the same type of resonant evolution that affects Crantor.
%
%
%
%

   \section{Discussion}
      The stability of the objects studied here reflects the chaos in the orbital evolution of the giant planets resulting from the 
      superposition of three-body mean motion resonances for Jupiter, Saturn and Uranus as identified by Murray \& Holman (2001). For large 
      objects, this translates into a chaotic but long-term stable system; for small bodies, this superposition turns out to be far from 
      stable. 
      \hfil\par
      The dynamical evolution of the previously documented Uranus' co-orbital 83982 Crantor (2002~GO$_{9}$) is very chaotic, the time-scale 
      necessary for two initially infinitesimally close trajectories associated with this object to separate significantly, or $e$-folding 
      time, is less than 1 kyr (de la Fuente Marcos \& de la Fuente Marcos 2013b). In contrast, the characteristic $e$-folding time of 
      2011~QF$_{99}$ during the current Trojan episode has been found to be nearly 10 kyr. Therefore, simulations over long time-scales 
      (e.g. $\gg$ 1 Myr) are not particularly useful in this case and we restrict our figures to just 2 Myr, maximum. Since the orbit of the 
      asteroid is chaotic, its true phase-space trajectory will diverge exponentially from that obtained in our calculations. However, the 
      evolution of most of the control orbits studied exhibit very similar secular behaviour of the orbital elements in the time interval 
      (-100, 100) kyr. It is worth noting that, for most 2011~QF$_{99}$ control orbits, the close encounters with Uranus happen well beyond 
      the ends of that interval. These encounters have the largest overall impact on the evolution of 2011~QF$_{99}$. The predicted future 
      evolution seems to be more consistent across control orbits than that of the past; i.e. the detailed dynamical evolution of this 
      object is less predictable into the past even if it may have been more stable then. There is a wide dispersion regarding the actual 
      time of insertion into Uranus' co-orbital region. The dynamical evolution of 2011~QF$_{99}$ as described by our integrations can be 
      considered reliable within the time interval indicated above but beyond that, we regard our results as an indication of the probable 
      dynamical behaviour of the minor body. However, the current dynamical status of the object is firmly established even if the details 
      of its dynamical evolution until it is ejected from the co-orbital region are affected by its inherently chaotic behaviour. Some 
      control orbits only show Trojan behaviour with a very short horseshoe episode during insertion and just prior to ejection. More often, 
      multiple transitions between the various co-orbital states are observed before ejection. 
      \hfil\par
      The comparative evolution of Uranus' co-orbital Crantor and candidates 1999~HD$_{12}$, 2002~VG$_{131}$ and 2010 EU$_{65}$ is displayed 
      in Fig. \ref{others}. It clearly shows that objects moving in 2010~EU$_{65}$-like orbits are far more stable than any other currently 
      known Uranus' co-orbital configuration, including that of 2011~QF$_{99}$. This is the result of its low eccentricity. Unfortunately, 
      2010~EU$_{65}$ has not been reobserved since 2010 June. Hypothetical objects moving in low-eccentricity orbits and locked in a Kozai 
      resonance (Kozai 1962) may be even more stable as in the case of the dynamically cold Kozai resonators described by Michel \& Thomas 
      (1996) or de la Fuente Marcos \& de la Fuente Marcos (2013a). These have not been found yet in the case of Uranus but they may exist 
      at other inclinations. The orbital evolutions of Crantor and 2002~VG$_{131}$ are remarkably similar and they briefly exhibit 
      Kozai-like dynamics, the argument of the perihelion librates (or remains almost constant) around 90\degr (see Fig. \ref{others}, panel
      F, second and third columns). 2002~VG$_{131}$ has not been reobserved since 2002 December. Both 2002~VG$_{131}$ and 2010~EU$_{65}$ 
      are very interesting targets for recovery; their short-term stability should make that task easier. In sharp contrast, the paths of 
      objects moving in 1999~HD$_{12}$-like orbits are so chaotic that the prospects of recovery for this object are rather slim; its last 
      observation was recorded almost 15 years ago and this time-span is practically of the same order as its $e$-folding time. 
      \hfil\par
\begin{figure*}
       \centering
        \includegraphics[width=\linewidth]{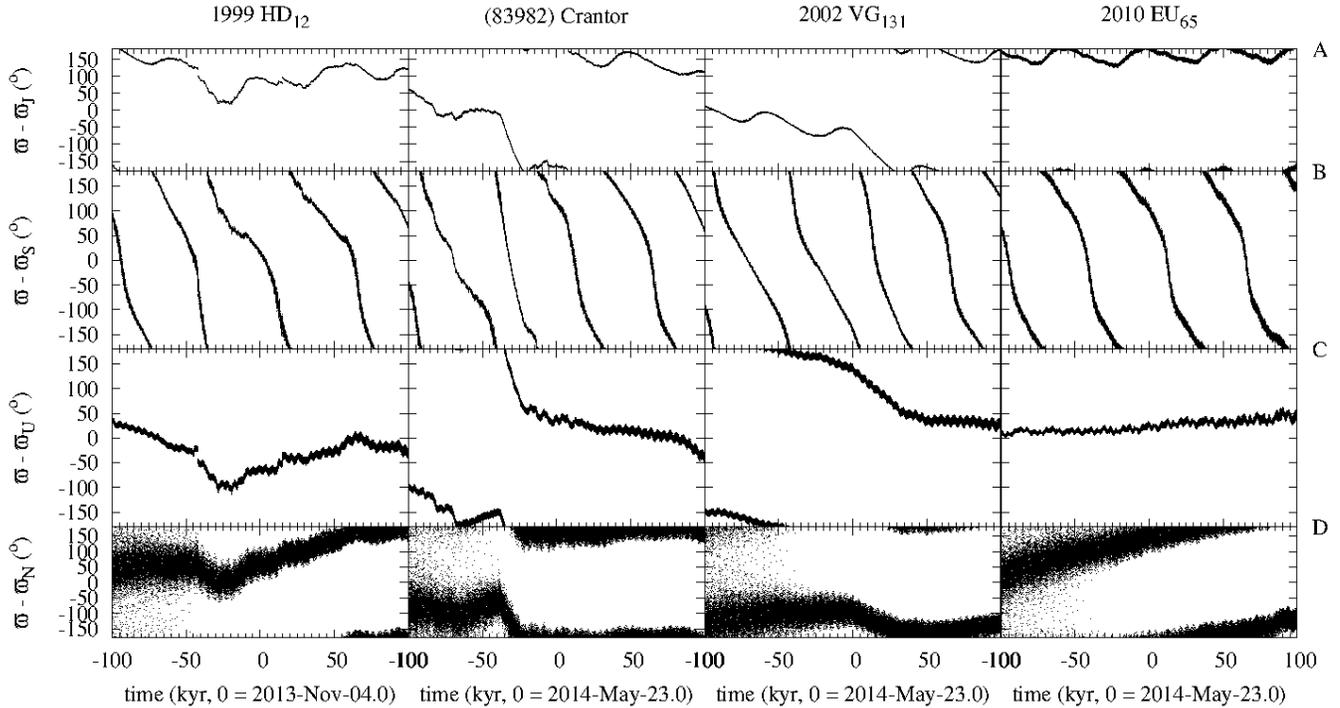}
        \caption{Time evolution of the relative longitude of the perihelion, $\Delta \varpi$, of 1999~HD$_{12}$, 83982 Crantor 
                 (2002~GO$_{9}$), 2002~VG$_{131}$ and 2010~EU$_{65}$. This figure is analogue to Fig. \ref{rellon}. Only in the case 
                 of Saturn, the resonant argument circulates.
                }
        \label{rellonO}
     \end{figure*}

      The destabilizing role of the 1:7 mean motion resonance with Jupiter in the case of Uranus' co-orbitals is just another example of how
      asymmetric libration in the region exterior to a perturbing mass is intrinsically more chaotic than that in the interior region. 
      Winter \& Murray (1997) studied the stability of asymmetric periodic orbits associated with the 1:$n$ resonances and how chaos is 
      induced at these resonances: hardly any actual asteroids are trapped in exterior 1:$n$ resonances in the Solar system. The role of the 
      exterior resonance with Jupiter on the dynamics of 2011~QF$_{99}$ cannot be neglected, nor can we ignore its influence on the dynamics 
      of other Uranian co-orbitals. As for secular resonances, if we represent the relative longitude of the perihelion, $\Delta \varpi$, 
      for all these objects, as we did for 2011~QF$_{99}$, we obtain Fig. \ref{rellonO}. Consistently with the case of 2011~QF$_{99}$, the 
      resonant argument circulates only for Saturn. With the exception of 2010~EU$_{65}$, the overall behaviour is similar to the one 
      observed and discussed for 2011~QF$_{99}$. 
      \hfil\par
%
%
%
%
      Of the 257 asteroids and 165 comets with semimajor axis in the range (6, 34) au currently in the MPC Database, five asteroids have 
      been identified as temporary Uranian co-orbitals (or candidates). Comet C/1942EA (V\"ais\"al\"a 2) is also a candidate to be a 
      temporary Uranian co-orbital. It means that nearly 2 per cent of the known asteroids with semimajor axis in that range are Uranian 
      co-orbitals. According to Alexandersen et al. (2013b), the intrinsic fraction should be 0.4 per cent with less than a factor of 2 
      variation. The current tally already appears to be five times larger than their theoretical expectations and this suggests that our 
      current understanding of the origin and dynamics of this resonant population may not be complete. Discrepancies may have their origin 
      in collisional processes. Three objects (Crantor, 2010~EU$_{65}$ and 2011~QF$_{99}$) have absolute magnitude $< 10$, the others have 
      $> 11$ mag. Using the results of collisional evolution calculations, Fraser (2009) has predicted the existence of a break in the size
      distribution (therefore, also in the absolute magnitude distribution) of TNOs somewhere in the 10--100 km range. Asteroids 
      1999~HD$_{12}$ and 2002~VG$_{131}$ may be fragments of larger objects, the result of collisional evolution. 
      \hfil\par
      Predictions from simulations (see Section 1) appear to suggest that additional Uranus' transient co-orbitals should exist, although 
      recent modelling by Alexandersen et al. (2013b) indicates that the current fraction is consistent with expectations. However, this
      appears to be in fact at odds with the observational evidence if all five co-orbitals and candidates are considered. On the other 
      hand, additional temporary (yet to be detected) co-orbitals may be in the form of intrinsically fainter objects (apparent magnitude at 
      perigee $>$ 23 mag, see Table \ref{discovery}), the result of collisional processes. If this hypothesis is correct, where can they be 
      preferentially found if their orbital elements are similar to those of the already identified co-orbitals and candidates? We try to
      answer this question in the following section. 

   \section{Hunting for Uranian co-orbitals: a practical guide}
      So far, the information collected from the various co-orbitals and candidates appears fragmentary, disparate and not particularly 
      useful to search for additional Uranus' co-orbitals. However, we already know a few robust facts about their orbits: i) their relative 
      (to Uranus) semimajor axes are likely $\leq$ 0.15 au, ii) their eccentricities are probably $\leq$ 0.5, and iii) there are perhaps 
      certain islands of stability in inclination. On the other hand, and in order to maximize the chances of discovery, we know that 
      objects must be observed near perigee (that is sometimes but not always near perihelion due to the relative geometry of the orbits). 
      The condition of being located at relatively small angular separations from Uranus is only true for quasi-satellites. It is clear from 
      our previous calculations that, in general, any angular separation from Uranus is possible, particularly for horseshoe librators. 
      Transient Uranus' co-orbitals easily switch between resonant states and they can change from Trojan to horseshoe librator on a 
      time-scale of kyr. If they must be found near perigee, when their apparent magnitude is the lowest because their distance from the 
      Earth is the shortest, the Solar elongation becomes a non-issue as it will always be $>$ 100\degr. 
      \hfil\par
      We have put all these ingredients at work in a Monte Carlo-type simulation, using the optimal ranges for semimajor axis, eccentricity 
      and inclination, and assuming that $\Omega$ and $\omega \in$ (0, 360)\degr, to compute the geocentric equatorial coordinates ($\alpha$, 
      $\delta$) of hypothetical objects moving in Uranus' co-orbital candidate orbits when they are closest to the Earth (i.e. when their 
      apparent magnitudes are the lowest). A uniform distribution is used to generate the orbital elements because the actual distribution
      in orbital parameter space is unknown. We do not assume any specific size (absolute magnitude) distribution as we are not trying to
      calculate any detection efficiency. However, any Uranian co-orbitals resulting from collisional evolution (i.e. objects like 
      1999~HD$_{12}$ or 2002~VG$_{131}$) with sizes below $\sim$10 km (or absolute magnitude $>$10 mag) reach perigee with apparent
      magnitude $>$22--23. 
      \hfil\par
\begin{figure}
        \centering
         \includegraphics[width=\linewidth]{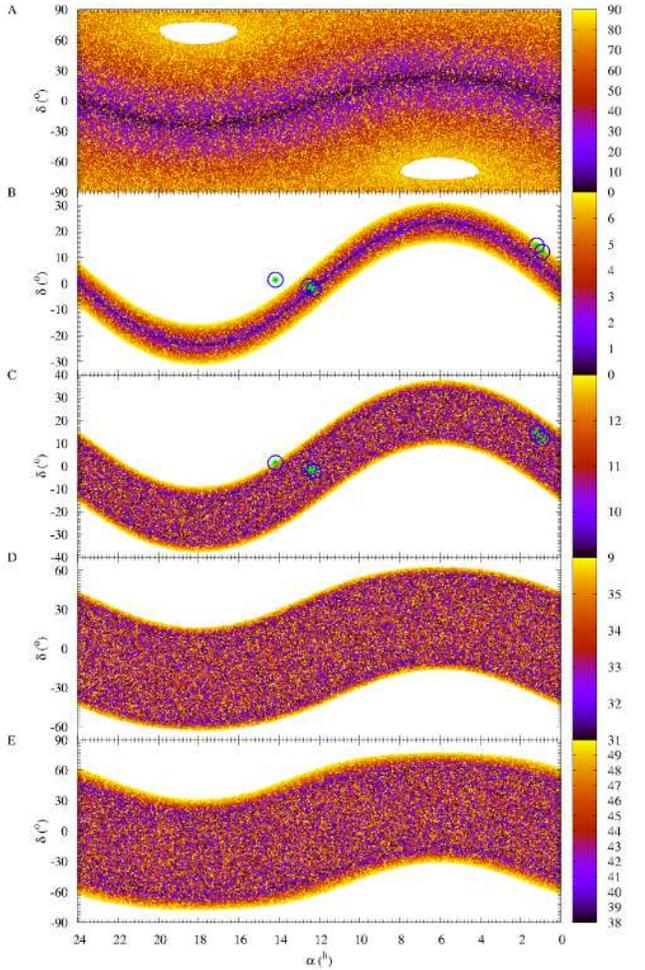}
         \caption{Distribution in equatorial coordinates of Uranus' co-orbital candidate orbits at perigee as a function of the inclination 
                  (regions as defined in Dvorak et al. 2010). Full set of prograde orbits (panel A); region A (panel B); region B 
                  (panel C); region C (panel D); region D (panel E). The results for several millions of test orbits are 
                  displayed. The green points in panels B and C represent the discovery coordinates of the five objects discussed in this 
                  paper (see actual data in Table \ref{discovery}). The darkest lane outlines the ecliptic. 
                 }
         \label{hunt1}
      \end{figure}
\begin{figure}
        \centering
         \includegraphics[width=\linewidth]{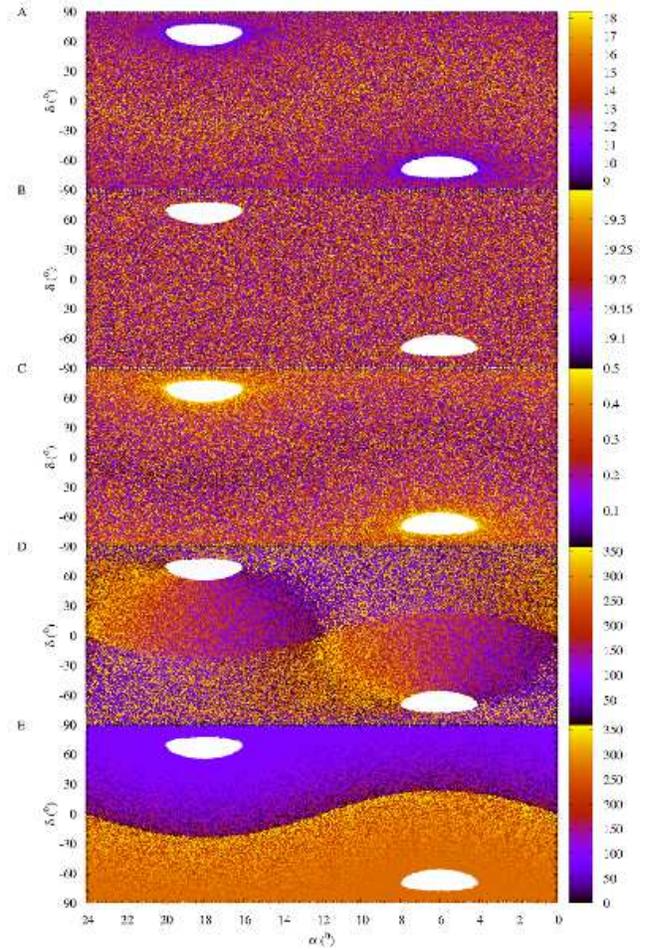}
         \caption{Distribution in equatorial coordinates of Uranus' co-orbital candidate orbits at perigee as a function of various 
                  orbital elements and parameters. As a function of the perigee of the candidate (panel A); as a function of $a$ 
                  (panel B); as a function of $e$ (panel C); as a function of $\Omega$ (panel D); as a function of $\omega$ 
                  (panel E).  
                 }
         \label{hunt2}
      \end{figure}

      Fig. \ref{hunt1} summarizes our findings in terms of location on the sky in equatorial coordinates at perigee as seen from the centre 
      of the Earth for $i \in$ (0, 90)\degr (panel A) and the islands of stability (panels B, C, D and E) in Dvorak et al. (2010). In this, 
      and in Fig. \ref{hunt2}, the value of the parameter in the appropriate units is colour coded following the scale printed on the 
      associated colour box (grey-scale in the printed version of the journal). Fig. \ref{hunt2} presents our results in terms of the 
      perigee and the orbital elements. The equatorial coordinates at the time of discovery of the five objects discussed in this paper are 
      displayed in Fig. \ref{hunt1}, panels B and C. Out of five discoveries, four of them appear in two pairs projected towards the same 
      two areas of sky (see Table \ref{discovery}). The fifth one is 83982 Crantor (2002~GO$_{9}$) itself that was serendipitously found by 
      the Near-Earth Asteroid Tracking (NEAT) program on 2002 April 12 at Palomar Observatory. The NEAT program was not designed to target 
      minor bodies in the outer Solar system. 
      \hfil\par
\begin{figure}
        \centering
         \includegraphics[width=\linewidth]{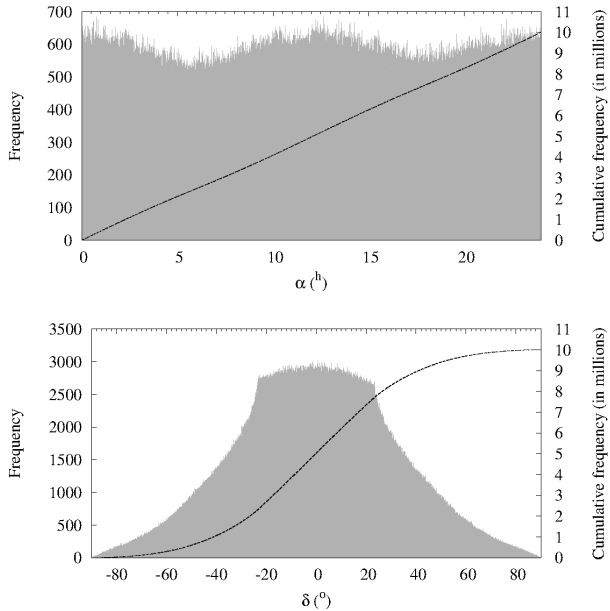}
         \caption{Frequency distribution in equatorial coordinates (right ascension, top panel, and declination, bottom panel) of 
                  Uranus' co-orbital candidate orbits at perigee. The best areas to search for these objects are located around 
                  right ascension 0$^{\rm h}$ and 12$^{\rm h}$ and declination $\in$(-20, 20)\degr.
                 }
         \label{radec}
      \end{figure}

      Statistically speaking, and assuming uniformly distributed orbital elements within the ranges pointed out above, the best areas to 
      search for these objects are located around right ascension 0$^{\rm h}$ and 12$^{\rm h}$ and declination $\in$(-20, 20)\degr (see 
      Fig. \ref{radec}). Hypothetical high-inclination, high-eccentricity objects are expected to be discovered at higher declinations. The
      frequency variation between observing at 6$^{\rm h}$ or 18$^{\rm h}$ and observing at 0$^{\rm h}$ or 12$^{\rm h}$ is higher than 20 
      per cent, 
      so statistically significant. Four out of five objects have been discovered at right ascensions near the statistically optimal values 
      of 0$^{\rm h}$ and 12$^{\rm h}$. On the other hand, the shortest perigees are preferentially found towards the ecliptic poles (see 
      Fig. \ref{hunt2}, top panel); the easiest to spot objects, if they do exist, have perigees well away from the ecliptic plane and they
      move in near polar orbits. In principle, these hypothetical objects may be unstable Kozai; but they could be submitted to the nodal 
      libration mechanism (see Verrier \& Evans 2008, 2009; Farago \& Laskar 2010) and that may make them long-term stable. It is obvious 
      that a significant fraction of the celestial sphere has not yet been surveyed for Uranian co-orbitals and that the scarcity pointed 
      out in Section 1 may not only be due to strong perturbations by the other giant planets but also the result of observational bias. The 
      most numerous objects may be the ones with absolute magnitude $\sim$14 mag (diameter $\sim$1 km, Fraser 2009) and they reach perigee 
      with apparent magnitudes 24--25.
%
%
%
%
%
%
      \begin{table}
        \centering
        \fontsize{8}{11pt}\selectfont
        \tabcolsep 0.35truecm
        \caption{Equatorial coordinates and apparent magnitudes (with filter) at discovery time for the five objects discussed 
                 in this paper. (J2000.0 ecliptic and equinox. Source: MPC Database.)
                }
        \begin{tabular}{lccc}
          \hline
             Object             & $\alpha$ ($^{\rm h}$:$^{\rm m}$:$^{\rm s}$) & $\delta$ (\degr:\arcmin:\arcsec) & $m$ (mag) \\
          \hline
             1999~HD$_{12}$     & 12:31:54.80                                 & -01:03:07.9                      & 22.9 (R)  \\
             (83982) Crantor    & 14:10:43.80                                 & +01:24:45.5                      & 19.2 (R)  \\
             2002~VG$_{131}$    & 00:54:57.98                                 & +12:07:52.4                      & 22.5 (R)  \\
             2010~EU$_{65}$     & 12:15:58.608                                & -02:07:16.66                     & 21.2 (R)  \\
             2011~QF$_{99}$     & 01:57:34.729                                & +14:35:44.64                     & 22.8 (r)  \\
          \hline
        \end{tabular}
        \label{discovery}
      \end{table}
%
%
%
%
%
%
%
%
%
%

   \section{Conclusions}
      In this paper, we have presented a detailed analysis of the orbital evolution and stability of present-day Uranus' co-orbitals, 
      focusing on how they got captured in the first place and what makes them dynamically unstable. Our calculations show that these 
      objects are submitted to multiple mean motion resonances and exhibit significant secular dynamics characterized by a complex structure 
      that sometimes includes apsidal corotations. The dynamical behaviour analysed here is a typical example of highly nonlinear dynamics, 
      resonances inside a resonance or even resonances inside a resonance inside a resonance. This is best studied using numerical 
      simulations not analytical or semi-analytical work. Some of our results verify predictions made by Marzari et al. (2003) after 
      studying the diffusion rate of Uranian Trojans.
      \hfil\par
      We confirm that 2011~QF$_{99}$ currently moves inside Uranus' co-orbital region on a tadpole orbit. The motion of this object is 
      primarily driven by the influence of the Sun and Uranus, although both Jupiter and Neptune play a significant role in destabilizing 
      its orbit. The resonant influence of Jupiter and Neptune was determinant in its capture as Uranus' co-orbital. The precession of the 
      nodes of 2011~QF$_{99}$, which appears to be controlled by the combined action of Saturn and Jupiter, marks its evolution and 
      short-term stability. A three-body mean motion resonance is responsible for both its injection into Uranus' co-orbital region and its 
      ejection from that region. The object will remain as Uranus Trojan for (very likely) less than 1 Myr. Even if 2011~QF$_{99}$ is one of 
      the most stable of the known bodies currently trapped in the 1:1 commensurability with Uranus, it is unlikely to be a primordial 1:1 
      librator.  
      \hfil\par
      Our comparative study of currently known Uranus' co-orbitals and candidates shows that the candidate 2010~EU$_{65}$ is more stable 
      than 2011~QF$_{99}$ because of its lower eccentricity (0.05 versus 0.18) even if 2010~EU$_{65}$ has higher orbital inclination (14\fdg8 
      versus 10\fdg80). On the other hand, a new candidate, 2002~VG$_{131}$, is found to exhibit dynamical behaviour very similar to the one 
      discussed for 83982 Crantor (2002~GO$_{9}$) in de la Fuente Marcos \& de la Fuente Marcos (2013b) and here. This new candidate is the 
      first identified quasi-satellite of Uranus; Crantor will also become a quasi-satellite in the near future. The presence of two objects 
      characterized by similar dynamics and found by chance suggests that others, yet to be discovered, may share their properties. Asteroid 
      1999~HD$_{12}$ may signal the edge of Uranus' co-orbital region. In any case, all these objects have present-day dynamical ages much 
      shorter than that of the Solar system; therefore, they are not members of a hypothetical population of primordial objects trapped in a 
      1:1 mean motion resonance with Uranus since the formation of the Solar system. They may be former primordial Neptune co-orbitals 
      (Horner \& Lykawka 2010) though.
      \hfil\par
      Horner \& Evans (2006) argued that present-day Uranus cannot efficiently trap objects in the 1:1 commensurability even for short 
      periods of time. However, the available evidence confirms that, contrary to this view and in spite of the destabilizing role of the 
      other giant planets, Uranus still can actively capture temporary co-orbitals, even for millions of years. Regarding the issue of 
      stability, both Crantor's and 2011~QF$_{99}$'s orbital inclinations are within one of the stability islands identified by Dvorak et al. 
      (2010) for the case of Trojans. Both candidates 2002~VG$_{131}$ and 2010 EU$_{65}$ move outside the stability islands proposed in that 
      study although they are not Trojans. The existence of 1999~HD$_{12}$ shows that not only inclination but also eccentricity play an 
      important role in the long-term stability of Uranus' co-orbitals. A larger sample of Uranus' co-orbitals is necessary to understand 
      better the complex subject of the stability of these objects although temporary Uranian co-orbitals are often submitted to complicated 
      multibody ephemeral mean motion resonances that trigger the switching between the various resonant co-orbital states, making them 
      dynamically unstable. 
      \hfil\par
      Currently available evidence suggests that the small number of known transient Uranus' co-orbitals may have its origin in 
      observational bias rather than in the strength of the gravitational perturbations by the other giant planets. Taking this into 
      account, the number of yet undiscovered transient Uranian co-orbitals may likely be as high as that of the Neptunian ones. Our results 
      can easily be applied to implement improved strategies for discovering additional Uranian co-orbitals. 

   \section*{Acknowledgements}
      We would like to thank the anonymous referee for his/her quick and to-the-point reports, and to S. J. Aarseth for providing one of the 
      codes used in this research. This work was partially supported by the Spanish `Comunidad de Madrid' under grant CAM S2009/ESP-1496. We 
      thank M. J. Fern\'andez-Figueroa, M. Rego Fern\'andez and the Department of Astrophysics of the Universidad Complutense de Madrid 
      (UCM) for providing computing facilities. Most of the calculations and part of the data analysis were completed on the `Servidor 
      Central de C\'alculo' of the UCM and we thank S. Cano Als\'ua for his help during this stage. In preparation of this paper, we made 
      use of the NASA Astrophysics Data System, the ASTRO-PH e-print server and the MPC data server.

\end{document}